# Unusual behavior of cuprates explained by heterogeneous charge localization

D. Pelc,[1,2] P. Popčević,[3,4] G. Yu,[2] M. Požek,[1*] M. Greven,[2*] N. Barišić[1,2,3*]

[1]Department of Physics, Faculty of Science, University of Zagreb, Bijenička cesta 32, HR-10000, Zagreb, Croatia

[2]School of Physics and Astronomy, University of Minnesota, Minneapolis, MN 55455, USA

[3]Institute of Solid State Physics, TU Wien, 1040 Vienna, Austria

[4]Institute of Physics, Bijenička cesta 46, HR-10000, Zagreb, Croatia

*Correspondence to: mpozek@phy.hr, greven@umn.edu, neven.barisic@tuwien.ac.at

**The cuprate high-temperature superconductors are among the most intensively studied materials, yet essential questions regarding their principal phases and the transitions between them remain unanswered. Generally thought of as doped charge-transfer insulators, these complex lamellar oxides exhibit pseudogap, strange-metal, superconducting and Fermi-liquid behaviour with increasing hole-dopant concentration. Here we propose a simple inhomogeneous Mott-like (de)localization model wherein exactly one hole per copper-oxygen unit is gradually delocalized with increasing doping and temperature. The model is percolative in nature, with parameters that are experimentally constrained. It comprehensively captures pivotal unconventional experimental results, including the temperature and doping dependence of the pseudogap phenomenon, the strange-metal linear temperature dependence of the planar resistivity, and the doping dependence of the superfluid density. The success and simplicity of our model greatly demystify the cuprate phase diagram and point to a local superconducting pairing mechanism involving the (de)localized hole.**

The parent (undoped) cuprate compounds are antiferromagnetic charge-transfer insulators[1] that evolve into conductors and superconductors upon doping. The superconducting transition temperature is highest at an optimal doping level of $p \approx 0.16$ holes per planar $CuO_2$ unit, and therefore the phase diagram is divided into underdoped and overdoped regions. The highly overdoped materials are conventional Fermi-liquid metals, with a large Fermi surface that corresponds to $1 + p$ holes per $CuO_2$ unit and a planar resistivity that exhibits quadratic temperature dependence[2,3], $\rho \propto T^2$. The underdoped region features the pseudogap[2], whose characteristic temperature $T^*$ decreases linearly with doping and extrapolates to zero at $p_c \approx 0.20$. It is well known that the resistivity exhibits unusual $\rho \propto T$ behavior in the "strange-metal" regime at temperatures above $T^*$, which has motivated a wealth of theoretical models. Furthermore, it was recently demonstrated[4,5,6] that Fermi-liquid behavior prevails below the characteristic temperature $T^{**}$ ($T^{**} < T^*$), with $\rho \propto T^2/p$.

In contrast to the resistivity and other observables, the transport scattering rate (determined from the cotangent of the Hall angle, $\cot(\Theta_H)$, also referred to as inverse Hall mobility) behaves remarkably simply: it exhibits quadratic temperature dependence across the phase diagram, with a prefactor that is essentially independent of doping and compound[7,8]. This universal behavior persists across $T^{**}$ and $T^*$, connects the two Fermi-liquid regimes at high and low doping with the intermediate strange-metal regime, and thus constitutes a crucial constraint for any model of the cuprates. To the best of our knowledge, this remarkable universality has not been captured theoretically. However, interpreted in the simplest possible manner, the transport data imply that the carrier density acquires temperature dependence above $T^{**}$ and, conversely, that the pseudogap formation and localization of exactly one carrier per $CuO_2$ unit are complete[8] below $T^{**}$.

Here we provide a simple explanation of the cuprate phase diagram by respecting the experimental observation of universal Fermi-liquid behavior, and by combining it with (de)localization-induced changes of itinerant carrier density. The (de)localization is essential to understand not only the "normal-state" at temperatures above $T_c$, but also key aspects of the superconducting state that are at odds with existing theories. Just like $T_c$, the zero-temperature superfluid density, $\rho_{s0}$ – the density of superconducting electrons divided by the effective electron mass at $T = 0$ – follows a dome-like doping dependence[9]. For translationally-invariant systems, such as pure Bardeen-Cooper-Schrieffer (BCS) superconductors, Leggett's theorem states that $\rho_{s0}$ corresponds to the density of mobile normal-state charge carriers[10]. Thus, $\rho_{s0}$ would be expected to increase with $p$, yet a detailed recent study found a monotonic decrease all the way up to $p_{c2} \approx 0.26$, the doping level where superconductivity disappears. The data cannot be explained with conventional 'dirty' BCS theory[9], where pair-breaking impurities break translational symmetry and decrease $\rho_{s0}$ (and $T_c$). Since superconductivity on the overdoped side vanishes as the concentration of localized holes decreases, whereas the Fermi-liquid behavior persists, the pairing glue must be associated with the localized hole[7]. Based on this insight, we will show that the doping dependence of $\rho_{s0}$ directly follows from the evolution of the normal-state properties, without any additional free parameters.

*The model.* Guided by these robust experimental results, and by the knowledge that the cuprates are inherently inhomogeneous[11,12,13,14], we propose the following model that associates this inhomogeneity with gradual Mott-like (de)localization and correctly captures the doping and temperature dependences of the resistivity and Hall effect as well as the doping dependences of $T^*$, $T^{**}$ and $\rho_{s0}$. First, the $p$ doped holes are taken to be mobile and to exhibit a Fermi-liquid scattering rate even at low doping levels[4,7]. In contrast, the one localized hole per $CuO_2$ unit is separated from the Fermi level by a local gap. This doping-dependent gap derives from the charge-transfer gap of the parent insulators, is expected to be the highest gap scale of the system, and should be inhomogeneous if the underlying crystal structure is disordered. Experimental signatures of this gap scale[15] include broad features in tunneling and photoemission spectroscopy, a characteristic mid-infrared peak in optical spectroscopy, features in magnetic susceptibility and resistivity, etc. (see Supplementary Information). Consistent with these data, the simplest possibility is (i) to take the mean value of the local gap, $\Delta_p$, to decrease linearly with doping, $\Delta_p = \Delta_0 (1 - p/p_c)$, where $\Delta_0$ is a constant, and (ii) to take the width $\delta$ of the gap distribution (a measure of the inherent inhomogeneity) to be doping-independent (Fig. 1a). The effective density of mobile carriers then becomes

$$p_{eff}(p,T) = p + \int_{-\infty}^{0} g(\Delta, \Delta_p, \delta) d\Delta + \int_{0}^{\infty} g(\Delta, \Delta_p, \delta) e^{-\Delta/2kT} d\Delta \qquad (1)$$

where $g(\Delta, \Delta_p, \delta)$ denotes the doping-dependent gap distribution. The first integral includes in the Fermi sea those holes that are no longer localized at $T = 0$, which corresponds to the part of the distribution that crosses the Fermi level with increasing doping (Fig. 1a). Similarly, the second integral includes those holes that are thermally excited over the gap. Note that the first two terms of Eq. (1) give the density of itinerant holes at $T = 0$, that $p_{eff} = 1 + p$ at high temperature and/or doping, and that the density of localized holes is $p_{loc} = 1 + p - p_{eff}$.

Neglecting compound-specific Fermi-surface complications that can cause the failure of the effective-mass approximation, but not of the applicability of the Fermi-liquid concept[7,8], the resistivity and Hall constant (per $CuO_2$ unit) are taken to be $\rho = C_2 T^2/p_{eff}$ and $R_H = 1/ep_{eff}$, respectively, in agreement with experiment[7,8]. The normal-state transport coefficients are then easily calculated once the distribution $g$ is specified. Notably, the universal temperature dependence $\cot(\Theta_H) \propto C_2 T^2$ is embedded in these calculations[7], i.e., the experimentally established $C_2$ is used to fix the absolute value of $\rho$.

We model $\rho_{s0}$ by assuming that the pairing is mediated by the localized holes, leading to:

$$\rho_{s0} = \gamma \sigma_{dc}^{res} T_c \; p_{loc} = \rho_{s0}^H \; p_{loc} \qquad (2)$$

This expression consists of two parts: (i) the conventional dirty-BCS expression for mobile holes (Homes' law)[16,17], $\rho_{s0}^H = \gamma \sigma_{dc}^{res} T_c$, where $\gamma = 35.2$ cm/K·μΩ is a universal numerical constant, $T_c$ is taken as a measure of the superconducting gap as per BCS theory, and $\sigma_{dc}^{res}$ is the residual normal-state conductivity, a measure of (pair-breaking) disorder; (ii) $p_{loc}$, obtained directly from modeling the normal state; the simplest possible, linear dependence is assumed between $\rho_{s0}$ and $p_{loc}$. Importantly, Homes' law fails to give the correct dependence of $\rho_{s0}$ on doping (or $T_c$) for overdoped compounds. Note also that Homes' law breaks down in the limit of low levels of pair-breaking (point) disorder[16], which in fact some cuprates, such as $HgBa_2CuO_{4+\delta}$ (Hg1201), might approach.

*Modification of the model at low superfluid densities.* In a local pairing scenario with short coherence lengths and underlying spatial inhomogeneity, it should be expected that superconducting gap disorder also plays an important role in regimes where the superconductivity is not fully established. Indeed, it was recently shown (for $p < p_c$) that superconductivity appears in a percolative fashion upon cooling toward $T_c$[18,19,20]. The underlying inhomogeneity induces a superconducting gap distribution of nearly universal width $T_0 \sim 30$ K, which leads to percolative superconductivity and is consistent with the notion that percolation phenomena play out on multiple energy/temperature scales. Close to the critical doping levels $p_{c1} \approx 0.06$ and $p_{c2} \approx 0.26$ that define the $T = 0$ extent of the bulk superconducting state, we thus expect the simple relation Eq. (2) to be modified by a percolative term. We will show that this provides an excellent description of the superfluid density in the vicinity of $p_{c2}$.

*Results.* In order to calculate the transport coefficients and superfluid density, the gap distribution function must be specified. The results do not crucially depend on the shape of the distribution (see Supplementary Information), but they critically depend on its "distance" from the Fermi level. In Fig. 1, we show a generic calculation with the simplest possible distribution, a Gaussian with $p_c = 0.2$, $\Delta_0 = 4000$ K, and Gaussian width $\delta = 700$ K (Fig. 1a). The density of itinerant holes at $T = 0$ obtained from Eq. (1) is shown in Fig. 1b. Notably, $p_{eff}(T = 0)$ begins to deviate from $p$ around optimal doping and smoothly crosses over to $1 + p$ holes at high doping/temperature[4,7]. In order to obtain the phase diagram in a manner similar to well-established experimental observations[21], the temperature dependence of the resistivity curvature, $d^2\rho/dT^2 \cdot p_{eff}(T = 0)$, is plotted in Fig. 1c. All defining normal-state features are captured: the $T^2$ regime in the underdoped region that ends at $T^{**}$; the characteristic temperature $T^*$; an extended $T$-linear-like regime around optimal doping; and a smooth crossover to Fermi-liquid behavior on the overdoped side. Remarkably, this is obtained with three doping-independent and experimentally-constrained parameters of the gap distribution (see also Supplementary Information): $\delta$ is consistent with the widths of features seen in optical conductivity and STM[13,22], $p_c$ is roughly the doping level where $T^*$ extrapolates to zero, and $\Delta_0$ is consistent with the charge-transfer gap scale[15]. In order to obtain numerical values of the resistivity, we use the universal transport scattering-rate constant $(C_2)$[8].

Figure 2 demonstrates excellent quantitative agreement of the model with transport results for Hg1201 and $La_{2-x}Sr_xCuO_4$ (LSCO). Motivated by STM and NQR results, we employ a slightly different, skewed Gaussian gap distribution, although the generic Gaussian (Fig. 1) leads to similar results (see Supplementary Information). Figure 2a,b shows the resistivity and Hall constant of underdoped[7] Hg1201 along with the model results. The particular doping level ($p \approx 0.1$) was chosen because all characteristic features are clearly observed there, including $T^{**}$, $T^*$ and the linear-$T$-like resistivity regime. Figure 2c,d demonstrates that the doping and temperature dependence of the resistivity curvature of LSCO is nicely captured by our model. The model also captures the 'anomalous criticality' observed in the resistivity above optimal doping[23], the temperature dependence of the Hall constant at all doping levels[7], and the universal dependence of the linear and quadratic resistivity coefficients on doping[4] (see Supplementary Information). Moreover, the distribution width $\delta = 600$ K that gives the best description for Hg1201 (Fig. 2a,b) is 25% smaller than for LSCO (Fig. 2c), in accordance with the notion that Hg1201 is somewhat less inhomogeneous: the Cu nuclear quadrupole resonance linewidths are roughly 25% narrower in optimally-doped Hg1201[24] than in LSCO[25].

Once the gap distribution parameters are known for the normal state, the superfluid density follows – crucially, no additional free parameters are introduced (except for the description of the narrow region close to $p_{c2}$ – see below) In order to calculate $\rho_{s0}$ for LSCO, we approximate $\sigma_{dc}^{res}$ in Eq. (2) by a small constant in the overdoped regime ($1/\sigma_{dc}^{res} = 15$ μΩcm, consistent with experiment[21,23]), use doping-dependent values from experiment below optimal doping[21], and calculate $p_{loc}$ from Eq. (1). As shown in Fig. 3, the result of this calculation is in excellent agreement with the experimentally-determined doping dependence of $\rho_{s0}$. On the overdoped side, $\rho_{s0}$ is limited by $p_{loc}$, whereas on the underdoped side, $p_{loc} = 1$ and Homes' law is recovered.

As noted, in the narrow regions at the edges of the superconducting dome (for $T_c \lessapprox T_0/2 \sim 15$ K), Eq. (2) ought to be modified by a percolative correction as a result of the intrinsic superconducting gap disorder. We take this corrective term from previous work on granular superconductors (see Methods) and use the detailed measurements[9] of $\rho_{s0}(p)$ in overdoped LSCO to test this idea. As seen from Fig. 3a, we again find excellent quantitative agreement with experiment: the model captures the kink at $T_c \sim 12$ K and, in particular, the percolative low-$T_c$ regime characterized by superconducting percolation scaling, $\rho_{s0} \sim T_c^{1.6}$ (Fig. 3b). The width of the superconducting gap distribution, $T_0$, is introduced as a free parameter (full width at half maximum), and the data are best fit with $T_0 = 23 \pm 1$ K, remarkably close to the value $27 \pm 2$ K obtained in previous studies of superconducting pre-pairing as a function of temperature[18,20]. Signatures of granular superconductivity have also been observed in experiments on underdoped thin films[26] of $YBa_2Cu_3O_y$, mirroring the percolative regime discussed here.

*Discussion.* Our phenomenological model captures both the normal-state and superconducting-state behavior at a *quantitative* level as a result of the experimentally-constrained input, yet it provides neither the microscopic origin of the inhomogeneous gap nor the exact nature of the pairing glue. Nevertheless, the model provides crucial insight into several salient aspects of cuprate physics – the origin of the pseudogap and related unconventional magnetism, the universal intrinsic disorder, the nature of the 'strange metal' state, and superconductivity – which we briefly discuss in what follows. Significantly, the myriad pseudogap features and the superconducting glue must be related to the intra-unit-cell physics of the localized hole. Our simple model does not explicitly include short-range inter-unit-cell correlations (which are probably important for the hole localization) and should thus be viewed as coarse-grained.

Within the proposed picture, the density of itinerant carriers at the Fermi level starts to decrease upon cooling at temperatures comparable to the localization gap scale, which corresponds to the opening of local (pseudo)gaps; these temperatures are very high in underdoped compounds, but approach zero at high doping. Once a significant fraction of the carriers is localized, the pseudogap is manifested in spectroscopic experiments, with a characteristic energy/temperature somewhat dependent on the experimental probe[15]. Antiferromagnetic fluctuations should accompany the appearance of localized holes, since the localization is a residue of Mott physics and of the associated antiferromagnetism. Indeed, neutron scattering experiments show a distinct decrease of antiferromagnetic fluctuations[27,28] above $T^*$ and a vanishing of low-energy antiferromagnetic fluctuations typically associated with local moments[29] above $p \sim 0.28$. Moreover, the unconventional $q = 0$ magnetic order[30,31] that has been found to occur below $T^*$ might also originate from fluctuations of localized holes (and thus preserve lattice translational symmetry), through a coupling to phonons[32], formation of coherent loop currents[31], or higher-order multipole effects[33]. Importantly, Kerr-effect measurements indicate broken time-reversal symmetry for $T < T^*$, yet show a memory effect[34] that stems from temperatures far above $T^*$. This can be attributed to the broad distribution of localization energies, with incipient order forming at high temperatures and $T^*$ being a lower, emergent scale. In fact, there should exist a higher emergent temperature scale $T_p$ ($T_p > T^* > T^{**}$) at which the localized holes percolate. This percolation scale may have been observed in susceptibility and resistivity measurements[15] on LSCO (see Supplementary Information).

An important feature of our model is universal gap disorder, which might either originate from electronic correlations[35] or structural effects[11,12], with a significant coupling between the two. Indeed inhomogeneity is an intrinsic feature of the cuprates and, in fact, perovskites are generically prone to it[11,13]. In the case of the cuprates, this is well documented through structural data[11] and various direct local probes, in particular STM and NQR[13,25,36]. It is a distinct possibility that mechanical strain and martensitic strain accommodation are its primary cause[12,11], which would result in prominent electronic features, since charge and spin degrees of freedom naturally couple to strain (and vice versa). For example, in the case of the colossal-magnetoresistance manganites, strong coupling between elastic and electronic degrees of freedom is indeed understood to be the cause of the observed multiscale inhomogeneity[37]. Additional evidence for inherent structural frustration in the cuprates includes conductivity and hydrostatic relaxation experiments that show stretched exponential behaviour characteristic of glassy materials[12] and X-ray experiments that find complex fractal interstitial-oxygen-dopant structures linked to percolative superconductivity[38]. We emphasize that ours is an effective low-energy model that packages in a simple fashion the complicated infrared (and higher energy) physics of the doped cuprates[39,40]. For example, the optical spectra are complex[39], and electronic correlations up to energies of several eV manifest themselves upon cooling below[40] $T_c$. Lattice anharmonicity effects may play a role and affect the relative importance of covalent and ionic characteristics[41].

In the proposed picture, two unconventional features – linear-$T$ resistivity and decreasing superfluid density – originate from the same underlying gap distribution. The linear-$T$-like resistivity appears whenever the temperature is high enough compared to a significant fraction of the gaps. This behavior is not tied to optimal doping and has been observed between $p \approx 0.05$ and $p_c \approx 0.20$, with[4,21,42] $\rho \propto T/p$. Close to optimal doping, the gap distribution extends down to the Fermi energy, and thus the linear-$T$-like resistivity also extends to low temperatures, again consistent with experiment[23]. As a result of the integration in Eq. (1), any featureless distribution will give a broad region of linear-like resistivity, up to temperatures determined by the distribution width ($\delta \sim 800$ K). This should be contrasted with power-law resistivity dependences due to quantum criticality, e.g., in heavy-fermion materials[43]. Although often conjured to explain cuprate properties, quantum criticality results in a number of scaling laws – e.g., for the Grüneisen ratio, the dynamical spin susceptibility, and the charge-carrier effective mass[43] – that have not been convincingly demonstrated, especially at the lowest temperatures/energies[23]. In contrast, the existence of a universal Fermi-liquid transport scattering rate and related scaling laws are well documented throughout the phase diagram[4,7,8]. Another possible source of linear-$T$ resistivity was suggested to be bad-metal incoherent transport[44], which would imply a short electronic mean-free path that violates the Mott-Ioffe-Regel limit. However, it was shown that the conventional, semi-classical Mott-Ioffe-Regel limit is a serious underestimate[4,45], and that the cuprates lie in the coherent regime, consistent with the absence of high-temperature resistivity saturation observed in most cases.

The doping dependence of the superfluid density is captured by the extremely simple Eq. (2). The well-known Uemura relation $\rho_{s0} \propto T_c$ for underdoped cuprates[46] follows directly from Eq. (2) if $\sigma_{dc}^{res}$ is doping-independent, which is rather well satisfied e.g., for YBCO[21,46]. In contrast, $\sigma_{dc}^{res}$ of underdoped LSCO exhibits considerable doping dependence[21]. The essential ingredients in the calculation of $\rho_{s0}$ are a local superconducting mechanism (as also suggested in ref. [9]) and an underlying spatially inhomogeneous pairing strength, which naturally explains the deviations from both Leggett's theorem and Homes' law for overdoped compounds.

While our model gives an overarching picture of the normal state and even captures the doping dependence of the superfluid density, it is less obvious how to treat the pairing mechanism (and thus $T_c$ itself). Nevertheless, our work opens the interesting possibility that the paring mechanism is similar to that first proposed by Little in the context of organic conductors[47]. This electron-electron mechanism involves (virtual) oscillations of localized charge that provide an interaction between itinerant carriers. As we have demonstrated, the cuprate phenomenology is consistently explained by assuming two-component physics: Fermi-liquid and localized. Thus it can be inferred that the dielectric fluctuations of the localized hole and its immediate environment are responsible for the effective pairing interaction and the $d$–wave nature of the superconductig order parameter (see also Supplementary Information). For such pairing, the relevant energy scale should be related to the localization gap, which decreases monotonically with doping, yet $T_c(p)$ is dome-shaped. This can be qualitatively understood by taking into account the experimental fact that the electron-electron interactions are not instantaneous, but retarded[48]: for any local superconducting glue with retarded interactions, two electrons/holes must be at the same location within a given timescale. This will not occur frequently in underdoped compounds where the carrier density is low, leading to a decrease in $T_c$ (see Supplementary Information). In overdoped compounds, the magnitude of the localization gaps decreases, which causes a concomitant decrease of $T_c$.

If inhomogeneous nanoscale hole localization is a generic property of perovskites, our model could be relevant to a wide class of doped charge-transfer or Mott insulators. A Fermi-liquid scattering rate, pseudogap effects, and nontrivial resistivity have been detected in titanates[49], whereas iridates show local gap disorder, Fermi arcs, and unconventional magnetism similar to the underdoped cuprates[50]. Most importantly, our model provides a unifying description of low-energy cuprate physics and captures the most relevant features of the phase diagram. All this simply follows from a spatially inhomogeneous hole (de)localization process, a simple Fermi-liquid behavior of the itinerant (delocalized) carriers, and a local superconducting mechanism associated with the localized holes.

**Methods.**

*Superfluid density in the percolation regime.* At doping levels close to the critical values $p_{c1}$ and $p_{c2}$ where bulk superconductivity disappears, we expect the underlying superconducting gap disorder (seen in temperature-dependent experiments at lower doping) to influence the superfluid density: when the gap distribution is close to zero energy, patches of superconducting material will form in the material at $T = 0$. Quantitatively, this effect can be included into the superfluid density by modifying Eq. (2):

$$\rho_{s0}^{corr} = \gamma \sigma_{dc}^{res} T_c p_{loc} f(T_c) \qquad (3)$$

where the function $f(T_c)$ is the gap distribution correction. The shape of $f(T_c)$ is taken from a previous study of a diluted granular superconductor[51], where it was found that the (normalized) superfluid density is equal to the normal-state conductivity of the equivalent percolating resistor network[52]. To a good approximation, the dependence of the superfluid density on the superconducting fraction (at zero temperature) is then $f = [(P - P_C)/(1 - P_C)]^\gamma$, where $P$ is the fraction of superconducting patches, $P_C$ the percolation threshold and $\gamma$ an exponent depending on the dimensionality of the percolation. Note that the function $f$ is normalized to the value at $P = 1$. In order to obtain the link between $P$ and the mean $T_c$ needed for Fig. 3a, we must specify the local gap distribution function (similar to the calculations of different responses in dependence on temperature[18-20]. $P$ is then the integral of the distribution function. Similar to previous work[18-20], we choose a Gaussian distribution with width $T_0$, leading to

$$f(T_c) = \left[ \frac{1}{2(1-P_C)} \left( 1 + E\left( 2\frac{T_c}{T_0} + E^{-1}(2P_C - 1) \right) - P_C \right) \right]^\gamma \qquad (4)$$

where $E$ and $E^{-1}$ denote the direct and inverse error function, respectively. The exponent depends on the dimensionality of the percolation process: for 3D, $\gamma \approx 1.6$, while for 2D, $\gamma \approx 1.0$ [52]. The data are clearly compatible with the 3D case (see inset of Fig. 3a), in agreement with temperature-dependent experiments probing superconducting percolation above $T_c$ in multiple cuprates[18,19]. The corresponding critical concentration is then[52] $P_C = 0.3$.

**Acknowledgments**
N.B. is grateful to the late S. Barišić for extensive discussions. D.P. and M.P. acknowledge funding by the Croatian Science Foundation under grant no. IP-11-2013-2729. P.P. acknowledges funding by the Croatian Science Foundation Project No IP- 2016-06-7258. The work at the University of Minnesota was funded by the Department of Energy through the University of Minnesota Center for Quantum Materials, under DE-SC-0016371 and DE-SC-0006858. The work at the TU Wien was supported by FWF project P27980-N36 and the European Research Council (ERC Consolidator Grant No 725521).

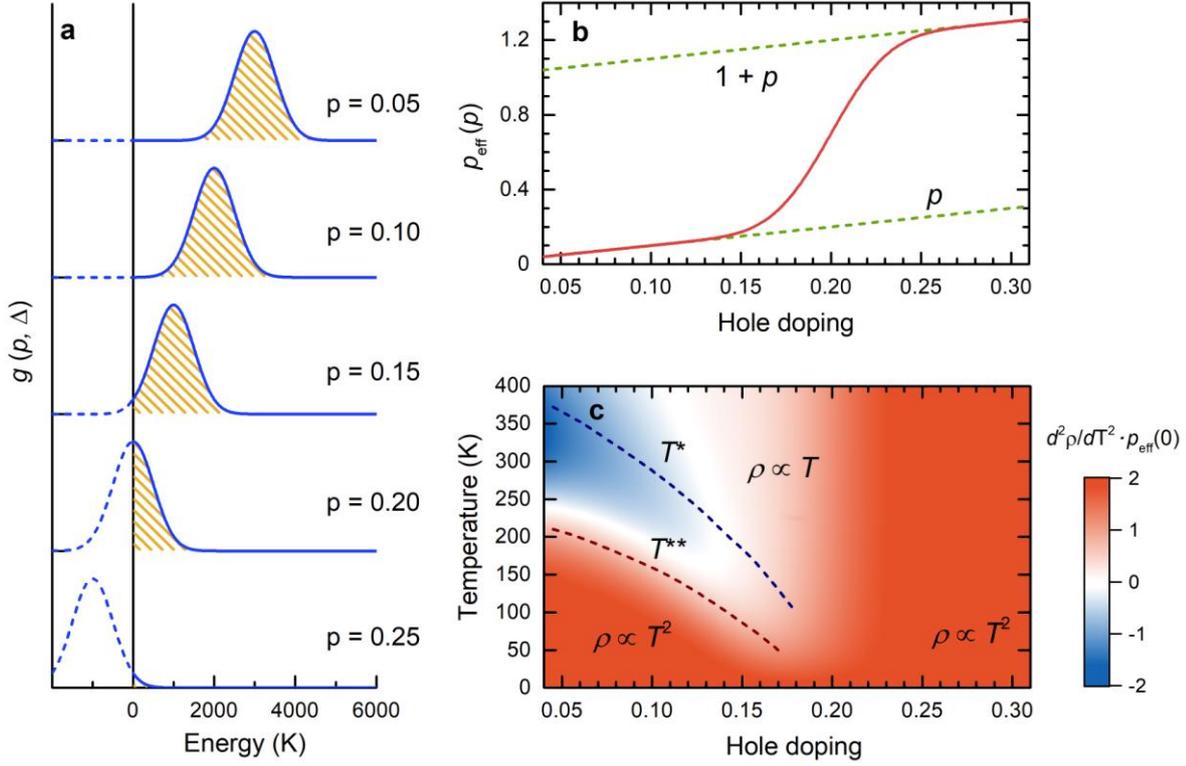

**Figure 1 | Gap inhomogeneity and phase diagram of the cuprates. a**, Gaussian gap distribution function $g$ at several doping levels, shown as a function of temperature. The parameters are $p_c = 0.2$, $\Delta_0 = 4000$ K and $\delta = 700$ K. The fraction of the distribution that has reached the Fermi level (the portion below energy $T = 0$ indicated with a dashed line) is added to the $p$ delocalized doped charge carriers at temperature $T = 0$. **b**, Effective density of delocalized carriers per $CuO_2$ unit at $T = 0$, obtained as the sum of the doped hole concentration $p$ and delocalized holes from the distributions shown in **a**. **c**, Second derivative of the resistivity in the absence of superconducting correlations, multiplied by $p_{eff}$ at $T = 0$. This result is obtained by combining the effective carrier density from Eq. (1), obtained with the gap distributions in **a**, with the experimentally-determined Fermi-liquid scattering rate of itinerant carriers[7]. The characteristic features of the phase diagram are apparent: a quadratic resistive regime at both low and high doping, the temperatures $T^{**}$ and $T^*$, and the linear-$T$-like regime around optimal doping.

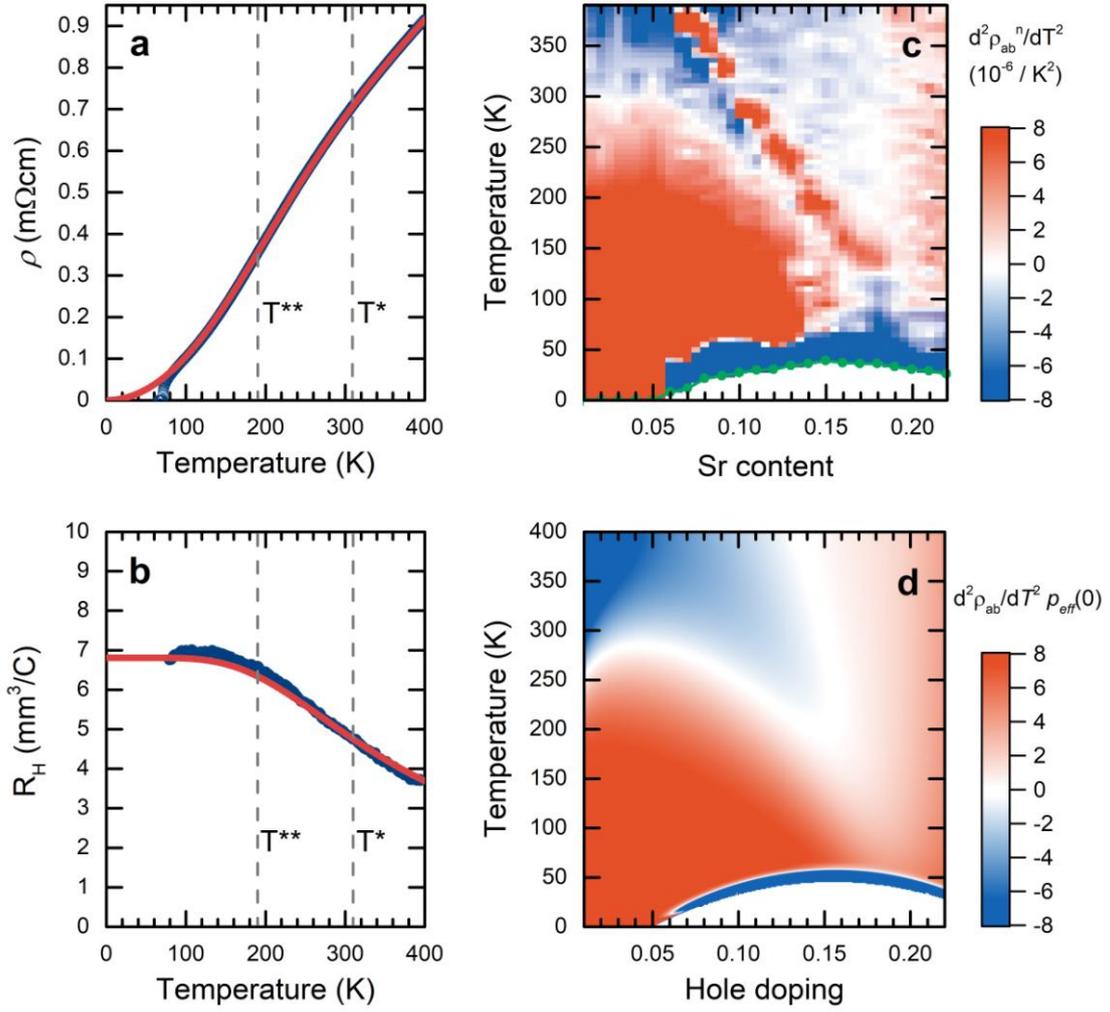

**Figure 2 | Comparison of the gap-distribution model to experiments. a**, Resistivity and **b**, Hall constant for underdoped Hg1201 ($p \approx 0.1$; blue symbols, data from ref. 7) compared to the gap disorder model (red lines). A skewed Gaussian gap distribution is used (see Supplementary Materials), with skew parameter $\alpha = 2$, and with $p_c = 0.2$, $\Delta_0 = 4000$ K and $\delta = 600$ K. The asymptotic value of the measured $R_H$ at $T = 0$ corresponds to about 90% of the nominal concentration $p = 0.10$ of mobile charge carriers, and the calculated curve is thus multiplied by the same factor. Note that this value is within the experimental uncertainty (due to sample size and shape uncertainty)[7]. **c**, Resistivity curvature map of LSCO (normalized at each doping level to $\rho(300 K)$) from ref. 21. In addition to $T^{**}$ and $T^*$, LSCO features a structural transition which is detected in the resistivity measurement (red downward-sloping line)[5]. **D,** Calculated resistivity curvature map, with added percolative pre-pairing regime[18] close to $T_c$. Resistivity is normalized by $p_{eff}(T=0)$, which is nearly equivalent to the experimental normalization, and multiplied by a factor of $4\cdot10^{-6}$ to have the same colour scale as in **c**. For LSCO the width of the gap distribution is $\delta = 800$ K, somewhat larger than for Hg1201. This is due to the expected modest difference in gap disorder[25,24] between Hg1201 and LSCO. All other model parameters are the same.

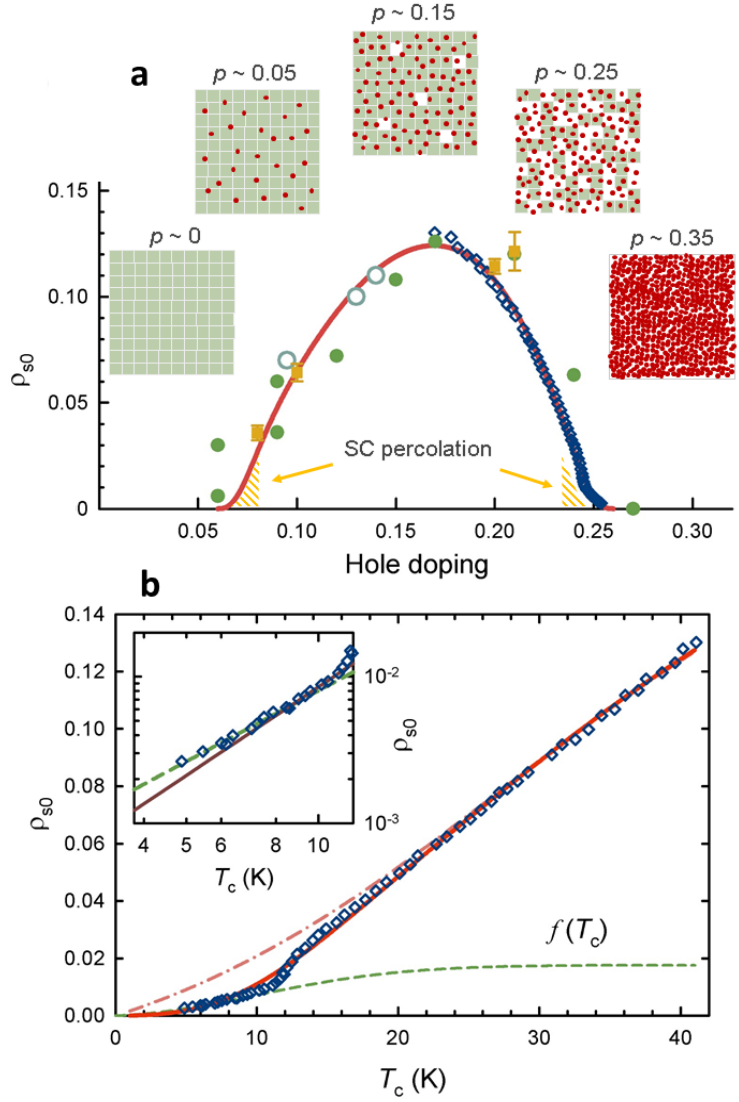

**Figure 3 | Superfluid density, inhomogeneous Mott localization, and superconducting percolation. a**, Zero-temperature superfluid density across the cuprate phase diagram calculated from Eq. (2) (full line). At low doping, $\rho_{s0}$ is limited by the density of mobile holes (red circles) and pair-breaking impurities, whereas at high doping the limiting factor is the density of localized holes (green squares). The insets are schematic, and one green square represents approximately four unit cells. A percolation regime is expected on both ends of the superconducting (SC) dome due to spatial inhomogeneity of the superconducting gaps[18,19]. Diamonds are data from ref. 9, whereas the other symbols are data for LSCO and for the closely related compound $La_{2-x}Ba_xCuO_4$ at $p = x = 0.095$, from optical conductivity (empty circles[53]), penetration depth (full circles[54]) and muon spin rotation (squares[46]). **b**, Superfluid density $\rho_{s0}$ of overdoped[9] LSCO in units of holes/unit cell (blue diamonds), modeled as a superposition of two gap inhomogeneity scales. Between optimal doping and $T_c \sim 12$ K ($0.19 < p < 0.24$), $\rho_{s0}$ is essentially proportional to the density of localized holes $p_{loc}$ (dot-dashed line), whereas at lower $T_c$ ($p > 0.24$) superconducting gap inhomogeneity causes percolation (dashed line – see Methods for the form of $f(T_c)$). The product of the two (full line) gives a reasonable description of the entire curve. Inset: log-log plot of $\rho_{s0}$ at low $T_c$ demonstrates good agreement with percolative scaling $\rho_{s0} \sim T_c^{1.6}$ (dashed line) compared to the previously used quadratic scaling[9] (full line). A power-law fit $\rho_{s0} \sim T_c^{\beta}$ below 9 K gives $\beta = 1.64 \pm 0.07$.

# Unusual behavior of cuprates

# explained by heterogeneous charge localization

D. Pelc,[1,2] P. Popčević,[3,4] G. Yu,[2] M. Požek,[1*] M. Greven,[2*] and N. Barišić[1,2,3*]

[1]Department of Physics, Faculty of Science, University of Zagreb, Bijenička cesta 32, HR-10000, Zagreb, Croatia

[2]School of Physics and Astronomy, University of Minnesota, Minneapolis, MN 55455, USA

[3]Institute of Solid State Physics, TU Wien, 1040 Vienna, Austria

[4]Institute of Physics, Bijenička cesta 46, HR-10000, Zagreb, Croatia

*Correspondence to: mpozek@phy.hr, greven@umn.edu, neven.barisic@tuwien.ac.at

**Content**

Supplementary Notes

1. Gap distribution
2. Phase diagram and skewness of distribution
3. Phase diagram and choice of doping dependences of distribution parameters
4. Hall coefficient and Hall angle
5. Resistivity coefficients
6. Dome-shaped doping dependence of $T_c$

Figures S1-S8

Supplementary References

**1. Gap distribution.** The cuprates are disordered on several levels, and different experimental probes access different types of disorder. In order to illustrate this salient point, we compile in Fig. S1 some results of local-probe measurements – scanning tunneling microscopy (STM) and nuclear quadrupole resonance (NQR) – for several representative compounds. STM studies of bismuth-based cuprates consistently show a broad distribution of pseudogaps and superconducting gaps[1], and a systematic analysis of different gap scales in $Bi_2Sr_2CaCu_2O_{8+\delta}$ (Bi2212) demonstrated the same underlying disorder on all scales and at all measured doping levels[2]. Copper NQR is a bulk local probe of net electrostatic disorder at the copper sites, which includes doping inhomogeneity, dopant-atom fields, structural distortions, etc. The significant NQR linewidths observed for both $La_{2-x}Sr_xCuO_4$ (LSCO) and $HgBa_2CuO_{4+\delta}$ (Hg1201) demonstrate the presence of intrinsic charge inhomogeneity[3,4] (note that the data for Hg1201 correspond to the quadrupolar satellite in NMR, and the frequencies are referenced to the line midpoint). In $YBa_2Cu_3O_{6+\delta}$ (YBCO), the NQR linewidths are smaller but still substantial (not shown in Fig. S1), despite the fact that this compound is highly crystallographically ordered for some oxygen doping levels[5].

Both electronic and structural origins of intrinsic disorder are possible, and there will always exist an interplay between the two due to electron-lattice coupling. Perovskites in general are prone to structural frustration leading to phase transitions or, if a long-range lattice distortion is avoided, to nano- and mesoscale disorder patterns[6,8]. Such nondiffusive cooperative structural displacements belong to a class of phenomena referred to as martensitic transitions, and are essentially unavoidable in lamellar systems such as the cuprates. The complex ensuing patterns of local unit cell deformations are a plausible candidate for the intrinsic disorder and gap distributions that underlie our phenomenological model. An interesting possibility is that the effects of structural disorder are exacerbated through coupling to strongly-correlated electrons.

Since transport measurements reveal that exactly one hole per $CuO_2$ unit, is (de)localized, it is natural to consider underlying Mott localization as the root mechanism, which is then modified by material complexities. In a simple one-band model without disorder, the (de)localization gaps the entire Fermi surface through a first-order phase transition. Notably, this process does not break any symmetry[7] leaving the Brillouin zone unchanged. However, understanding the evolution of such (de)localization upon doping, in a system with multiple bands (i.e., explicitly including the oxygen degrees of freedom) in the presence of structural/electronic disorder (that presumably causes gradual, inhomogenous (de)localization) creates enormous difficulties for theory. It is therefore not surprising that the 'infrared' physics of the cuprates is extremely complex and exhibits no clear/sharp features in optical conductivity, tunneling spectroscopy and photoemission. Our model packages these complexities in a simple effective manner, without purporting to provide microscopic insight.

The model features only three experimentally-constrained parameters associated with the gap distribution function: $\Delta_0$ (the extrapolated mean gap at zero doping, $\Delta_{p=0}$), $p_c$ (the doping level where the mean gap crosses the Fermi level), and $\delta$ (the width of the gap distribution). In order to calculate the magnitude of the resistivity, the universal scattering rate parameter $C_2$ known from experiment[21] is additionally required. In principle, the gap distribution parameters would be best evidenced from STM, a real-space local probe. Given that the cuprates are inherently inhomogeneous in the bulk[6,8], a surface probe like STM will naturally detect this inhomogeneity as well. An example of the fact that STM provides insight into bulk properties is the remarkably good agreement between STM and bulk X-ray experiments regarding the modulation wave vector of the subtle short-range charge-density-wave (CDW) order in the pseudogap state[9,10,11]. However, only a very limited number of cuprate compounds are accessible to STM, and the gap disorder is expected to be to some extent compound-dependent. STM data are therefore only used as a guide for our calculations, especially on the overdoped side, where there are virtually no relevant data. Extensive measurements on Bi2212 show that the largest local gap scale extracted from STM (denoted $\Delta_1$ in ref. 2) is about 800-1000 K around $p = 0.08$-$0.10$ and that it decreases monotonically with doping. This scale corresponds rather well to the intermediate scale $E_{up}$ of the three scales ($E_{hump} > E_{up} > E_{lp}$) compiled in ref. 12. On the other hand, our gap scale $\Delta_p$ is larger by about a factor of two and better corresponds to the largest scale $E_{hump}$ in ref. 12, which manifests itself in numerous probes[13-18], including optical conductivity[14,15], photoemission spectroscopy[16,17] and tunneling junction experiments[18,19] (Fig. S2b). Figure S2b is a compilation of the doping dependence of these latter results for a number of cuprates. The remarkably good agreement between $\Delta_p$ and $E_{hump}$ implies that the localization of the one hole per $CuO_2$ unit is the highest ranking process, whereas the other energy scales and related processes are emergent.

In order to obtain the generic values of $\Delta_0$, $p_c$ and $\delta$ used in Fig. 1, we simply assume a linear doping dependence of $\Delta_p$, with an extrapolated endpoint approximately equal to $p_c$ and a doping-independent distribution width (see Fig. S2b). Remarkably, in order to achieve detailed quantitative agreement between our model and experiments, these assumptions only need to be slightly relaxed (see below). The generic parameters for the Gaussian gap distribution in Fig. 1 are $\Delta_0$ = 4000 K, $\delta$ = 700 K (Gaussian width) and $p_c$ = 0.2. We note that we express all gap energies in units of Kelvin (with 1 meV ≈ 11.6 K), but that no assumptions related to the physical relationships between energy and temperature scales are made. In particular, in order to calculate transport properties via Eq. (1), only the gap energy values are needed (see also below).

For simplicity, we take $\delta$ to be a constant, in rough agreement with the broad features seen at the $E_{hump}$ scale in photoemission and tunneling experiments and with the scatter in the data in Fig. S2b. Moreover, our choice of $\delta$ is consistent with the STM results on Bi2212, where (lower-energy) gap distribution widths are only weakly doping dependent, and amount to about 500 K (Fig. S1a-c). Since the high-energy 'hump' scale is about a factor of two larger than the largest scale extracted from[2] STM, it is reasonable and in agreement with experimental observations that the width of different gaps scales with their mean gap values, to take the width to be about two times larger as well.

Importantly, our approach is compatible with optical conductivity experiments, which provide another way to estimate the gap distribution parameters. In addition to the coherent Fermi-liquid contribution at low frequencies[20], broad mid-infrared peaks are clearly resolved (see Fig. S2a) that shift to lower frequencies with increasing doping and eventually merge with the coherent response at high doping ($p > 0.20$)[14,15]. This feature was previously identified as a signature of the localized holes[21]. In Fig. S2b, we demonstrate that the mean energy of this peak (detailed doping dependence determined for both LSCO and YBCO[14,15]) indeed coincides with $E^*_{hump}$ and thus $\Delta_p$. In strongly underdoped LSCO ($p$ = 0.02), the width is about 1000 K and the mean energy is approximately 5000 K (see also Fig. S2a).

As shown in Fig. S3, an interesting observation can be made by plotting the temperature and doping dependence of the calculated density of localized carriers along with data for the highest available characteristic temperatures $T^*_{high}$ for LSCO ($T^*_{high} > T^* > T^{**}$) obtained from resistivity and magnetic susceptibility (from ref. 12, where the scale is denoted as $T^*_{hump}$). It can be seen that the data nicely follows the line corresponding to 1 − $p_{loc}$ ~ 20%, which might be an indication of the existence of a critical concentration and underlying percolation process.

**2. Phase diagram and skewness of distribution.** The gap distribution is usually skewed to the high-energy side[1,2,22] (see Fig. S1). In order to quantify and assess the impact of this tendency, we parametrize the distribution as a Gaussian multiplied by its integral:

$$g(\Delta, \Delta_p, \delta) = 2\phi(\Delta, \Delta_p, \delta) \int_{-\infty}^{\alpha\Delta} \phi(\Delta', \Delta_p, \delta) d\Delta' \qquad (1)$$

where α is a dimensionless parameter which controls the skewness of the distribution, and ϕ is a normalized Gaussian distribution with mean $\Delta_p$ and full width at half maximum δ. For α = 0, the distribution reduces to a simple Gaussian, and its skewness can be continuously varied by changing α.

In Fig. S3, we test different values of the skew parameter, as well as different functional forms for the gap distribution, to demonstrate that this yields essentially the same phase diagrams. The main features are rather insensitive to the choice of distribution shape, mainly due to the fact that the resistivity calculation involves an integral over energy. Along with skewed Gaussians, we tested a shifted gamma distribution that features a heavier tail at high energies and gives a broader linear-$T$ resistivity region on the overdoped side of the phase diagram.

**3. Phase diagram and choice of doping dependences of distribution parameters.** The simplest assumption of a linear decrease of the mean energy and doping-independent distribution width may be relaxed by introducing additional parameters and assumptions. While presently such assumptions are purely phenomenological (and detract somewhat from the main message of the simplest possible calculation), we expect a microscopic theory to be capable of providing the true doping dependences, at least in principle. In order to test the robustness of our calculation, we introduce a curvature into the dependence of the mean gap on doping, using a function of the form

$$\Delta_p(p) = \Delta_0 [1 - \tanh(p/p_c)/\tanh(1)] \qquad (2)$$

This function still crosses zero at $p = p_c$ (and contains no additional parameters), but has an upward curvature at higher doping.[23]. A similar function can be used for the dependence of the distribution width on doping, but with a more general form

$$\delta(p) = \delta_0 [1 - \beta \tanh(p/p_c)] \qquad (3)$$

where β is a numerical constant. The more constrained form with β = 1/tanh(1) cannot be used for δ, since it would lead to a zero distribution width at $p_c$ and nonphysical divergences in the calculations. We choose β ~ 0.4, but again it turns out that the exact value is not very important. The main features of the phase diagram are unchanged, but introducing these nontrivial doping dependences of $\Delta_p$ and δ broadens somewhat the region of the phase diagram where the resistivity is linear in temperature on the overdoped side (Fig. S5). In order to introduce also a curvature in $\Delta_p(p)$ at low doping, in line with some experiments (Fig. S2b), we further modify Eq. (2) and cast it in the form

$$\Delta_p(p) = \Delta_0 \left[ 1 - \left( \tanh(p/p_c)/\tanh(1) \right)^{1/2} \right] \quad (4)$$

without introducing additional free parameters. This form gives better agreement between calculated and measured Hall coefficients for strongly underdoped LSCO (see next Section), as well as a better match between the modeled mean energy (dashed line in Fig. S2b) and the characteristic high-energy scale from experiment. Moreover, it has the physically appealing feature of $\Delta_0$ being approximately 1 eV, the charge transfer gap at zero doping as determined from Hall-effect measurements[30]. Yet again it introduces no considerable changes in the overall picture, and thus in the main text we keep the simplest dependencies for $\Delta_p$ and δ.

**4. Hall coefficient and Hall angle.** Numerous interesting interpretations of the transport behaviour of the cuprates were proposed in the past. However, certain experimental facts were unknown until rather recently. In particular, only in recent years it was firmly established that the charge carriers exhibit universal Fermi-liquid behavior within the pseudogap regime[20,21,24,25] below a characteristic temperature $T^{**}$ ($T^{**} < T^*$): the planar resistivity scales as[24] $\rho \propto T^2/p$, the optical scattering rate exhibits Fermi-liquid temperature-frequency scaling[20], the magnetoresistivity obeys Kohler's rule (Fermi-liquid scaling)[25]. Moreover, the temperature dependence of the cotangent of the Hall angle (i.e., the inverse Hall mobility) is universally $\cot(\Theta_H) \propto C_2 T^2$, not just at temperatures below $T^{**}$, but throughout the whole accessible doping range[21]. These scaling results pose a very strong constraint, as they demonstrate that non-Fermi liquid models[26,27] are not appropriate in the pseudogap and 'strange metal' regimes.

At temperatures below $T^{**}$ (for underdoped compounds), the Hall constant is a rather accurate measure of the doped carrier density[21,30] $p$. It has been long known that the Hall constant exhibits strong temperature dependence above the pseudogap temperature[28],[29,30] which, when interpreted in the simplest possible manner, indicates that the effective carrier density acquires a temperature dependence. Although such behavior is not exhibited by textbook Landau Fermi liquids, it is consistent with a repopulation of the Fermi surface from arcs (for $T < T^{**}$) to a large $(1 + p)$ Fermi surface (at high temperatures and doping levels). Indeed, measurements of the Hall constant in LSCO up to 1000 K show a strong decrease, consistent with considerable delocalization of charge[30]. In fact, early work on LSCO already indicated approximate $1/T$ scaling at temperatures[28] above $T^*$, and there were considerations that included a temperature-dependent effective carrier density. In particular, ref. 31 describes a $T^2$ Hall mobility and temperature-dependent carrier density in thallium cuprates; ref. 32 discusses the possibility of a linear-$T$ resistivity arising from a $1/T$ carrier density temperature dependence, and in refs. 33 and 30 the effective carrier number was modeled by including an activation term with a single gap energy.

Yet as remarked in the main text, only recent results for the cotangent of the Hall angle show simple universal quadratic temperature dependence across $T^*$ and to high doping levels, which demonstrates that even in the strange-metal regime the underlying scattering rate is Fermi-liquid-like[21]. The states on the arcs (elongated with increasing temperature and doping) are therefore exactly the same as those found at high doping levels, where clearly a Fermi-liquid exists. In the present work, we therefore keep the experimentally established Fermi-liquid scattering rate throughout.

In addition to enabling a straightforward calculation of the doping and temperature dependences of the resistivity (apart from a non-universal, compound-specific "residual" resistivity contribution[34], our model gives the Hall constant simply as the inverse effective number of carriers $1/p_{eff}$. In Fig. S6, we compare our result to the data of Ref. 30. The temperature dependences are captured well, even in the optimally-doped region where single-gap fitting is not satisfactory[30]. However, as expected, the absolute magnitudes do not agree completely and require a multiplicative correction as a consequence of the change of Fermi-surface curvature with doping. In LSCO, the curvature decreases toward the antinodal region of the Brillouin zone that is increasingly relevant near optimal doping, which leads to an overall decrease of the Hall constant[21]. In the strongly underdoped compounds, even better agreement between the calculated Hall constant and experiment is obtained by using Eq. (4) for the dependence of the mean gap on doping, which possesses an upward curvature as doping tends to zero (which is not the case for the similar Eq. (2)).

**5. Resistivity coefficients.** Here we compare our model to two other pivotal experimental results. The relatively wide doping range in which the planar resistivity of optimally/overdoped LSCO exhibits linear-$T$ dependence was taken as a sign of 'anomalous criticality'[35]. Whereas this result cannot be easily understood with conventional theories of quantum criticality[35], it naturally follows from our model. Namely, the (de)localization gap disorder causes a rather smooth evolution from the underdoped Fermi liquid, with $p$ mobile carriers at low temperatures, to the overdoped Fermi liquid with $1 + p$ mobile carriers. As the gap structure crosses the Fermi level, an extended region of doping with appreciable linear-$T$ resistivity appears. For a quantitative comparison to the data, we employ the same analysis as in ref. 35 on the resistivity obtained from our model. Note that the model does not take into account the changing geometry of the LSCO Fermi surface with overdoping[36]. Following ref. 35, the model results are fit to a parallel resistor formula $1/\rho = 1/(A_0 + A_1 T + A_2 T^2) + 1/\rho_0$, where $A_0$, $A_1$ and $A_2$ are the resistivity coefficients, and $\rho_0$ a fixed limiting resistivity value corresponding to high-temperature resistivity saturation[35]. The temperature range of the fit is 5 to 200 K at all doping levels, which is arbitrary, but the same as in ref. 35. We again use Eq. (1) from the main text for the effective density of mobile carriers $p_{\text{eff}}$, and we calculate $\rho = C_2 T^2 / p_{\text{eff}}$. The constant $C_2$ captures the universal scattering rate (as deduced from Hall-angle measurements[21]) and determines the absolute value of $\rho$. The comparison of the calculated coefficients $A_1$ and $A_2$ with experiment in Fig. S7 demonstrates remarkable agreement considering the complex Fermi surface of LSCO and the simplicity of our model (e.g., the linear doping dependence of the mean gap $\Delta_p$ beyond optimal doping is probably overly simplistic; a relationship like Eq. (2) would extend the linear-$T$ region to higher doping levels).

We also establish that the maximum in $A_1$ observed experimentally around $p = 0.18$ is a consequence of the limited temperature range of the data and fit (up to 200 K); calculations for a significantly wider range smooth out the maximum (see below). If the upper temperature bound is decreased from 200 K, the maximum becomes even more prominent and its width decreases – this is expected, since the linear-$T$ term appears at low temperatures only close to the doping level $p_c$ where the mean of the gap distribution crosses the Fermi level. Note that $p_c$ derived from an extrapolation of $T^*$ is slightly smaller than our $p_c$ – this is due to the fact that the $T^*(p)$ line in the phase diagram corresponds approximately to 0.9 localized holes per unit cell, while the mean gap $\Delta_p$ corresponds to 0.5 localized holes. The $T^*$ line will thus cross zero at lower doping than the $\Delta_p$ line.

The second experimental study we discuss here is ref. 24. This work analyzes results for numerous cuprates for the planar resistivity above $T^*$ and below $T^{**}$ and concludes for underdoped compounds that the sheet resistance coefficients $A_{1\square}$ ($T$–linear behavior; fit to data for $T > T^*$) and $A_{2\square}$ (from $T^2$ Fermi-liquid fit for $T < T^{**}$) are universal and scale approximately as $1/p$. It was also found that, beyond optimal doping, once superconductivity is suppressed with a large $c$–axis magnetic field and the planar resistivity is best fit to a polynomial form rather than pure linear or quadratic behavior, $A_{1\square}$ decreases sharply whereas $A_{2\square}$ levels off (Fig. S8). We calculate the resistivity coefficients by fitting a polynomial temperature dependence in the low-temperature range (1 – 100 K) to extract $A_{2\square}$. To obtain $A_{1\square}$, we fit a linear function to the calculated resistivity in a temperature range $T^* < T < T^* + 200$ K (where the resistivity is predominantly linear) for underdoped compounds, and above ~0.12 hole doping we fit a parallel resistor formula (as described above) in the temperature range 100 – 400 K. The two procedures to extract $A_{1\square}$ merge smoothly around $p = 0.12$. The extracted model values of $A_{1\square}$ are somewhat sensitive to the fit procedure, as also exemplified by the maximum in Fig. S7. Therefore, $A_{1\square}$ had to be multiplied by a factor of 1.2 in order to best match the experimental values. Note also that photoemission experiments in overdoped LSCO find a Fermi surface with both electron-like and hole-like sections, corresponding to a net carrier number[36] of $1 - p$. For the sake of simplicity, we have neglected any Fermi surface changes and assumed a carrier density of $1 + p$ at high doping (as seen in other cuprates with simple Fermi surfaces[37]). This difference in carrier density may explain the above-mentioned factor 1.2 discrepancy. For the description of the superfluid density of ovedoped LSCO (see Fig. 3), these corrections are unimportant, since only $p_{\text{loc}}$ is needed and the contribution from itinerant carriers is described by Homes' law (see Eq. (2) in the main text).

As seen from Fig. S8, the experimental trends of $A_{1\square}(p)$ and $A_{2\square}(p)$ are well captured by our model. The $1/p$ dependences below optimal doping follow because for underdoped compositions, $p_{\text{eff}} = p$ in a relatively wide temperature region; the simple Fermi-liquid formula for the conductivity then yields $A_{2\square}$ directly, whereas $A_{1\square}$ is the lowest-order correction to the Fermi-liquid dependence in this doping range, and thus also scales with $1/p$. As optimal doping is approached, the complex gap structure begins to cross the Fermi level, and $p_{\text{eff}}$ departs from $p$ (as seen in Fig. 1b). Beyond optimal doping, $A_{1\square}$ strongly decreases as the overdoped Fermi-liquid regime is approached, whereas $A_{2\square} \propto 1/(1+p)$, which is only weakly $p$-dependent. Due to the relatively large temperature range of the fits to the result generated by our model (up to 400 K, compared to 200 K in Fig. S7a), $A_{1\square}$ no longer exhibits a maximum.

In order to best match the LSCO data, the coefficients in Figs. S7 and S8 are obtained with slightly different model parameters than the phase diagrams in Figs. 1, 2, S1 and S2: $\Delta_0$ = 3900 K, $\delta$ = 800 K, and $p_c$ = 0.22. In LSCO, the extrapolated pseudogap endpoint is indeed slightly shifted to higher doping compared to Hg1201 or YBCO, justifying the employed shift. Given the simplicity of our model, the agreement with experiment is remarkable.

Very recent work attributes the linear-$T$ resistivity to strong quasi-one-dimensional umklapp scattering of antinodal quasiparticles, which do not contribute to the Hall conductance that stems from the nodal (Fermi-liquid) parts of the Fermi surface[38]. It is not obvious how to reconcile this scenario with the experimental fact that $\cot(\Theta_H) \propto T^2$ across $T^{**}$ and $T^*$, and the success of our minimal model in describing many experimental results demonstrates that it is not necessary to invoke such a mechanism. Yet if such a scattering process does play a role, we emphasize that gap disorder is an essential ingredient required to understand the smooth crossover to the overdoped regime.

Our calculations demonstrate that an extended linear-$T$ regime in the resistivity can be obtained in a simple manner, without invoking bad-metal transport or quantum criticality. However, we note that a quantum critical point associated with the emergent unconventional magnetism below $T^*$ may well be present, but without a direct influence on transport and superconducting properties, especially at high temperatures.

**6. Dome-shaped doping dependence of $T_c$.** In conventional superconductors, the effective phonon-induced interaction between quasiparticles is not instantaneous, but occurs on a characteristic timescale related to the phonons involved in the pairing. This gives rise to a relation between the density of electrons and $T_c$, which we believe is also important in the case of the cuprates (regardless of the nature of the pairing mechanism). Within the well-known Eliashberg theory of strongly-coupled phononic superconductors, optic phonons with a characteristic energy $\omega_p$ are the source of the superconducting glue. The interaction is thus relatively confined in space and time, and two quasiparticles must therefore be within a characteristic volume for pairing to occur. The superconducting transition temperature was shown to be given by[39]

$$\ln \frac{T_c}{\omega_p} \approx -\frac{1+\lambda}{\lambda - \mu^* - \lambda\mu^*} \qquad (4)$$

where λ is the electron-phonon coupling and μ* is the Morel-Anderson Coulomb pseudopotential[40] which measures the repulsive electrostatic interaction. The coupling may be obtained, e.g., in a simple pseudopotential approximation (for a nearly-free electron metal), using the adiabatic approximation and effective electron-ion potential $V_q$, which depends on the wave-vector $q$. When the retarded nature of the interaction is taken into account, the coupling becomes[39]

$$\lambda = \frac{1}{r_s} \frac{\langle V^2 \rangle}{\langle \tilde{\omega}_p^2 \rangle} \qquad (5)$$

where $r_s$ is the mean radius of a sphere containing one quasiparticle, and the angular brackets denote averages over all wave vectors. For an isotropic material, $1/r_s$ is clearly proportional to $p^{1/3}$, where $p$ is the density of mobile charge carriers. Hence the effective coupling (and thus the $T_c$) depends on the electron density.

The above example of a phonon-induced retarded interaction showcases an important point that is not restricted to the particular pairing mechanism: a pairing interaction that is local in both space and time induces a dependence of $T_c$ on the density of mobile charge carriers. We have shown compelling evidence that the superconducting glue in the cuprates is closely related to the localized hole. Therefore, one would expect the localization scale $\Delta_p$ to be the relevant energy scale for pairing, equivalent to $\omega_p$ above. However, this is clearly not the only factor that affects the superconducting transition temperature, since otherwise $T_c(p)$ would simply monotonically decrease with increasing doping. The dome-shaped doping dependence of $T_c$ can nevertheless be understood quite generally, without knowing the exact nature of the pairing glue, if one takes into account the fact that the pairing interaction is local and retarded. Recent ultrafast pump-probe experiments have shown that the interaction is retarded, and that the characteristic timescale τ is on the order of 20 fs near optimal doping[41]. This value corresponds roughly to $1/\hbar\Delta_p$. According to our results, the pairing is local, and two mobile holes have to be within the pairing volume on a timescale comparable with τ. On the underdoped side, the mobile-hole density is low, the probability of such an event is diminished, and the pairing energy decreases. A nonzero hole concentration is then needed for pairing because of competing repulsive electrostatic interactions, in analogy with the well-known phonon case above. On the overdoped side, the density is high, but the characteristic energy $\Delta_p$ decreases, again lowering $T_c$. Thus arises the superconducting dome. In the case of the cuprates, the calculation of $T_c$ ought to be modified even if a local-phonon mechanism were responsible for superconductivity. The energies $\Delta_p$ are very high and approach the Fermi energy at low doping, which implies that the adiabatic approximation (or, equivalently, the Migdal theorem in phononic superconductivity[42]) is not satisfied. This is again a more general remark, not necessarily limited to a phonon-related glue. Yet we note that the strong-coupling theory involves optic phonons and an electron-phonon interaction that is local in space, and might thus be of more direct quantitative relevance to the cuprates than might seem at first sight.

Furthermore, since the characteristic energy is spatially inhomogeneous in the cuprates, the mobile holes might self-organize to increase the probability of coupling to the most favorable local sites. The resulting superconducting gap disorder is then detected via superconducting percolation[43,44] in the vicinity of $T_c$, but its relation to the distribution of $\Delta_p$ is likely nontrivial. Self-organization might be the reason why the width of the superconducting gap distribution $T_0$ is de facto universal[43,44], whereas the high-energy gap disorder may show some variation among cuprate families – e.g., the transport data for Hg1201 and LSCO are best described with gap distribution widths differing by about 25% (see main text). Yet the localization gap distributions are still remarkably similar. At low superfluid densities, the overall phase stiffness of the condensate is small and the phase may locally adapt to avoid regions of lower pairing energy. Such superconducting phase relaxation may play an important role[45], with possible modifications to our simple calculation.

Finally, in our model, the pairing glue is provided by the localized hole. As noted in the main text, a two-component electron picture may lead to pairing mediated by localized charge fluctuations. In this respect, several experiments give evidence of the importance of oxygen-oxygen charge transfer within the unit cell. The low temperature tetragonal (LTT) phase in lanthanum-based cuprates involves a small structural change which barely affects normal state transport, but breaks the symmetry of the in-plane oxygens and sharply decreases[46] $T_c$. A similar oxygen symmetry breaking and $T_c$ suppression occurs locally around Zn impurities in copper-oxygen planes, as shown by $^{67}$Zn NQR measurements[47]. The oxygen symmetry is also spontaneously broken by charge density waves[48] which are in competition with superconductivity. A fluctuating oxygen-copper-oxygen polarization could thus play an important role in providing the glue[49]. Experiments that directly probe the intra-unit-cell physics are needed to elucidate the exact superconducting mechanism as well as how exactly the hole gets localized.

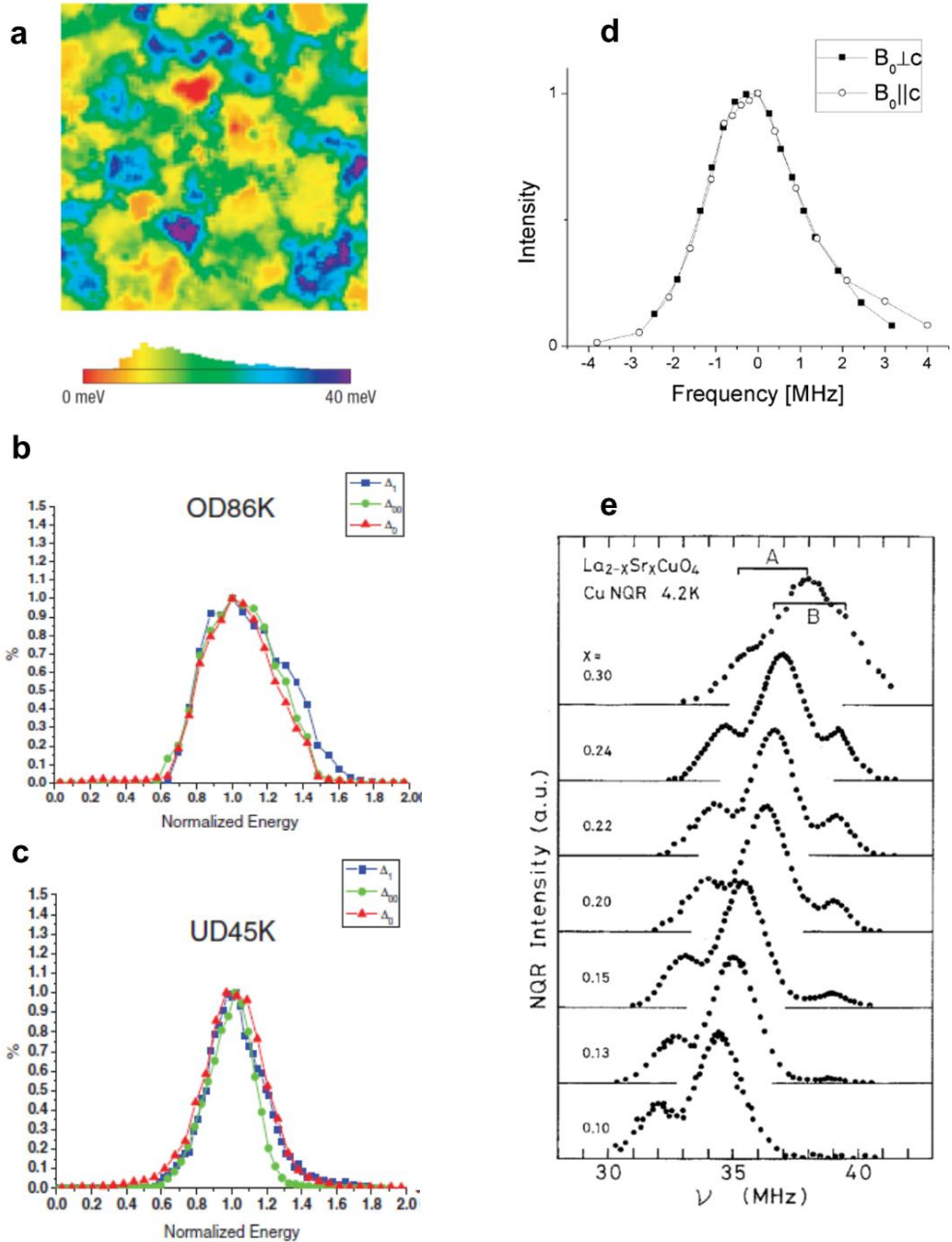

**Figure S1 | Local probes of disorder in cuprates. a,** High-energy gap map from an STM study of Bi2201 and corresponding histogram, in a field of view of 180x180 Å$^2$ (adapted from ref. 1). **b,** and **c,** are distributions of three fitted gap components from an STM study of Bi2212 for an overdoped and underdoped sample, respectively (adapted from ref. 2). Significantly, the same underlying inhomogeneity manifests itself in all gap scales. The energies are normalized to the mean values, but the distribution widths on an absolute energy scale are essentially the same. **d,** $^{63}$Cu NMR quadrupolar satellite line in optimally-doped Hg1201, where the linewidth is a measure of local electrostatic disorder (adapted from ref. 3). **e,** Doping dependence of nuclear quadrupole resonance (NQR) spectra of LSCO in a wide doping range (adapted from ref. 4), showing that the line-width is essentially unchanged across the phase diagram. Notably, widths of the $^{63}$Cu A lines are slightly larger than for Hg1201. The smaller feature below the $^{63}$Cu A line is the $^{65}$Cu A line, whereas the B line stems from sites close to the Sr impurities.

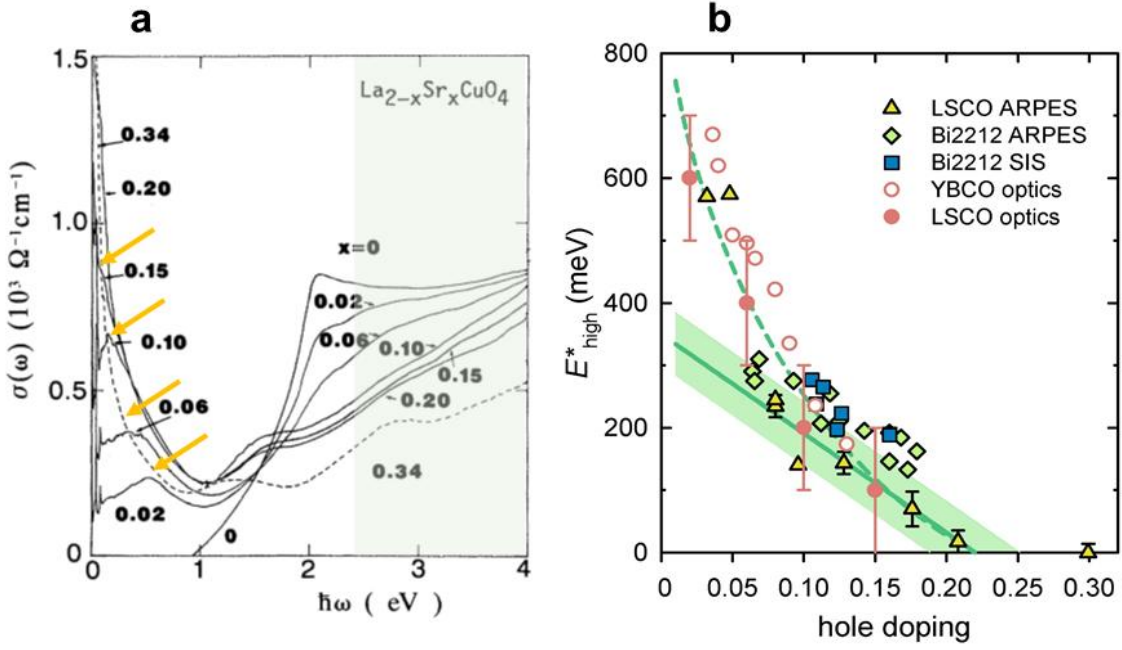

**Figure S2 | High-energy features in cuprates. a,** Optical conductivity of LSCO at 300 K, showing a mid-infrared feature at energies consistent with our mean delocalization gap (arrows indicate the characteristic feature that corresponds to the mean gap value) evolving from the charge transfer gap and merging into the coherent Drude peak at high doping. Adapted from ref. 14. **b,** Comparison of the highest-energy characteristic scale in different cuprates. The ARPES and SIS data correspond to the 'hump' scale, while the optical conductivity data are the energies of the mid-infrared peak. The solid green line is our parameterization of the localization gap for LSCO, whereas the shaded green band indicates the gap distribution width. The dashed line is an alternative parameterization, Eq. (4), which includes upward curvature and extrapolates to the charge transfer gap (~ 1 eV) at zero doping. ARPES and SIS data are adapted from ref. 12 while the data are from multiple experiments, see ref. 12 for original references. LSCO optical conductivity peak energies are from **a**, and YBCO optical conductivity is adapted from ref. 15.

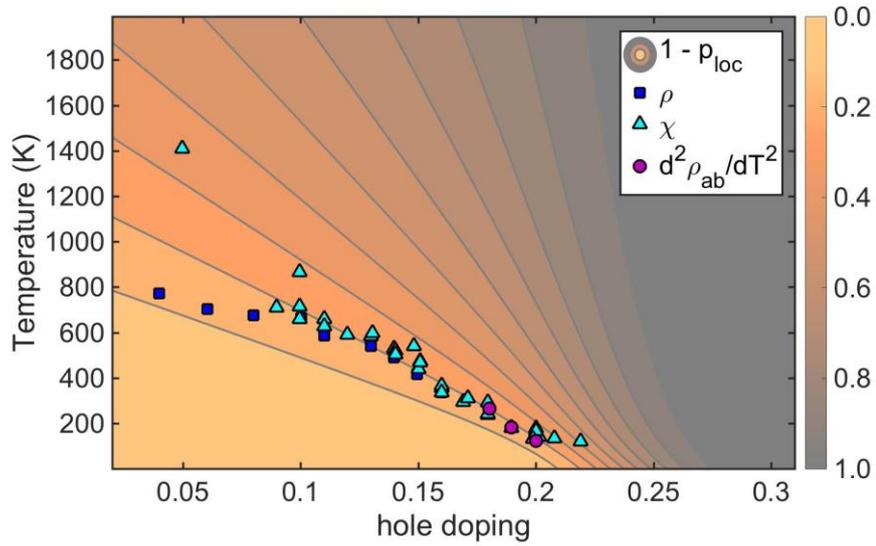

**Figure S3 | Characteristic temperature and localization.** Comparison for LSCO of the highest measured characteristic temperature scale $T^*_{high}$ from resistivity and magnetic susceptibility measurements with the calculated fraction of localized holes. The iso-lines on the contour plot correspond to increments of 10%. It is seen that, between $p \sim 0.10$ and $0.20$, the data closely follow the line for 20% delocalization. This may hint at an underlying percolation/connectivity transition. Data are from ref. 12 and references therein.

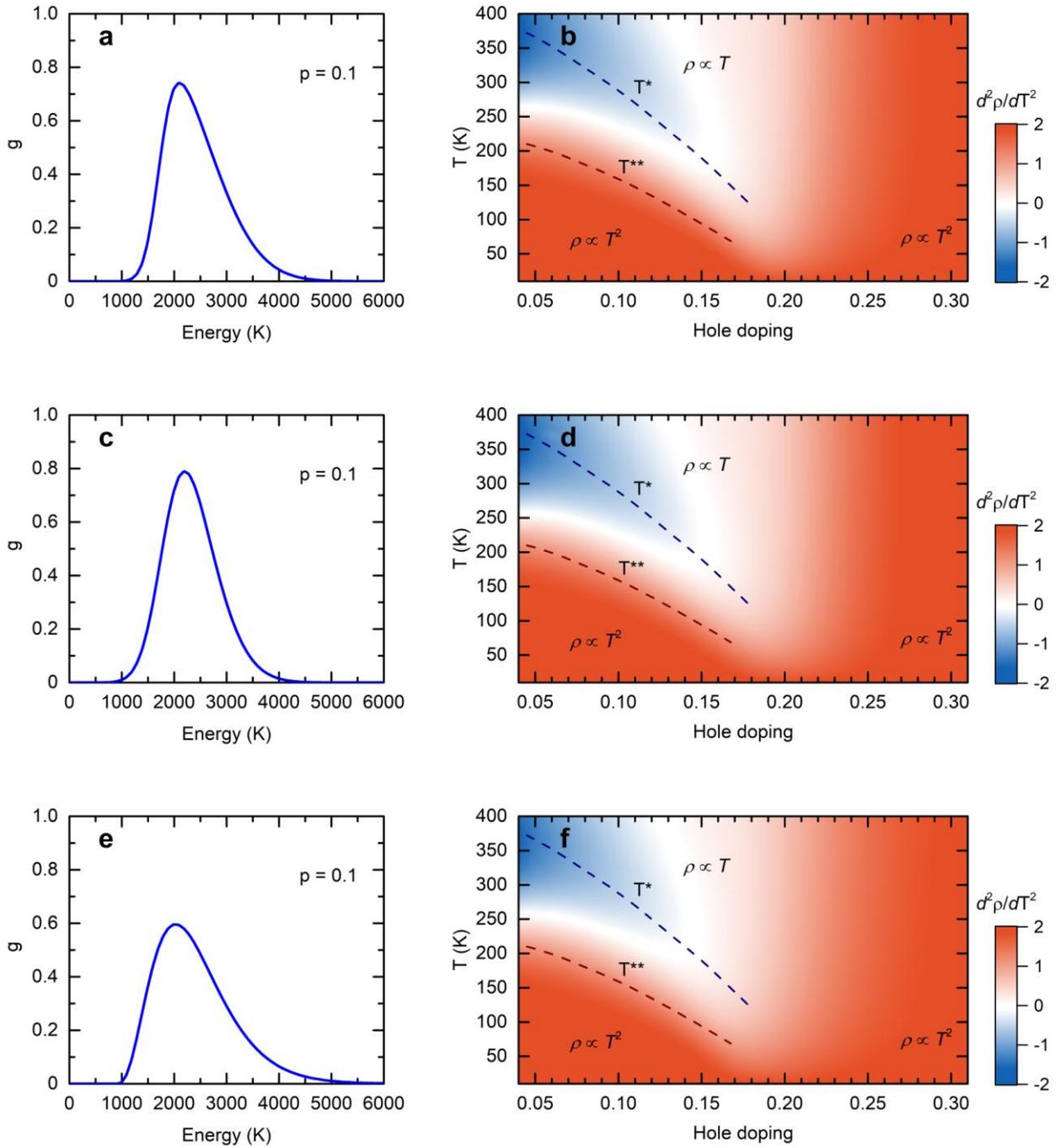

**Figure S4 | Normal state phase diagram for different gap distributions. a,** Skewed Gaussian distribution, with skew parameter α = 4 and width parameter δ = 1000 K. **b,** The corresponding resistivity curvature phase diagram. **c,** A less skewed Gaussian distribution, with skew parameter α = 1 and width parameter δ = 800 K. **d,** The corresponding phase diagram. **e,** A shifted gamma distribution, with exponent $l$ = 3 and width parameter δ = 400 K. **f,** The corresponding phase diagram. All distributions are shown at nominal doping level $p$ = 0.10, and a linear dependence of the mean gap energy on doping is assumed. The distributions are normalized to the area under the curves.

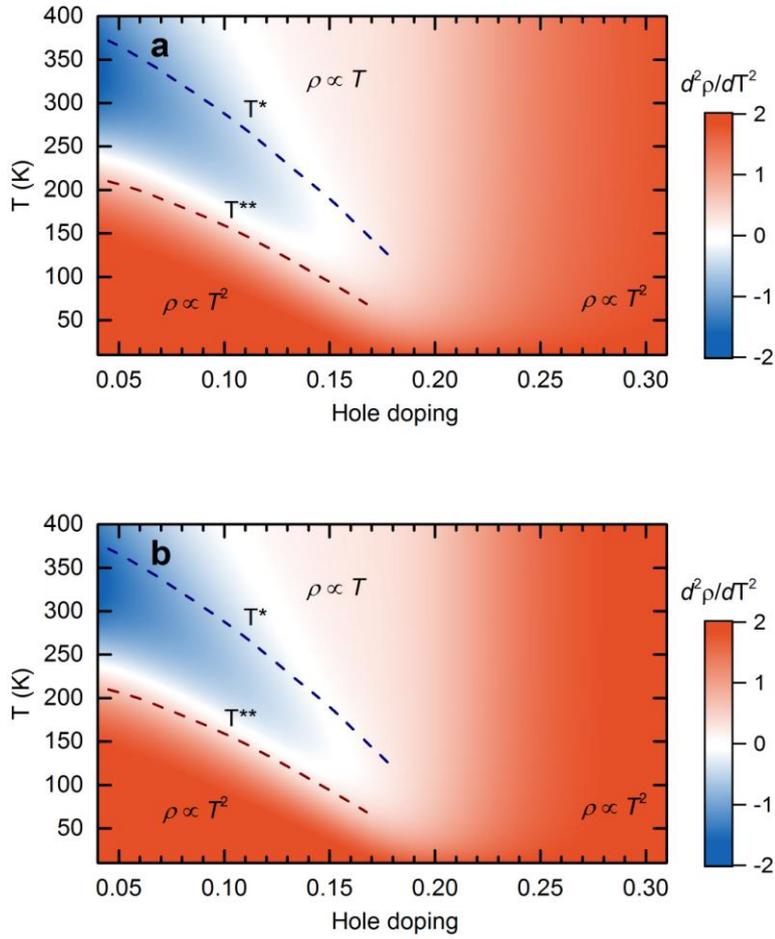

**Figure S5 | Normal-state phase diagrams for two doping dependences of the gap distribution parameters.** Hyperbolic tangent doping dependence of the distribution width and mean, with two values of the width doping dependence parameter (see supplementary text): **a,** $\beta = 0.4$, and **b,** $\beta = 0.6$.

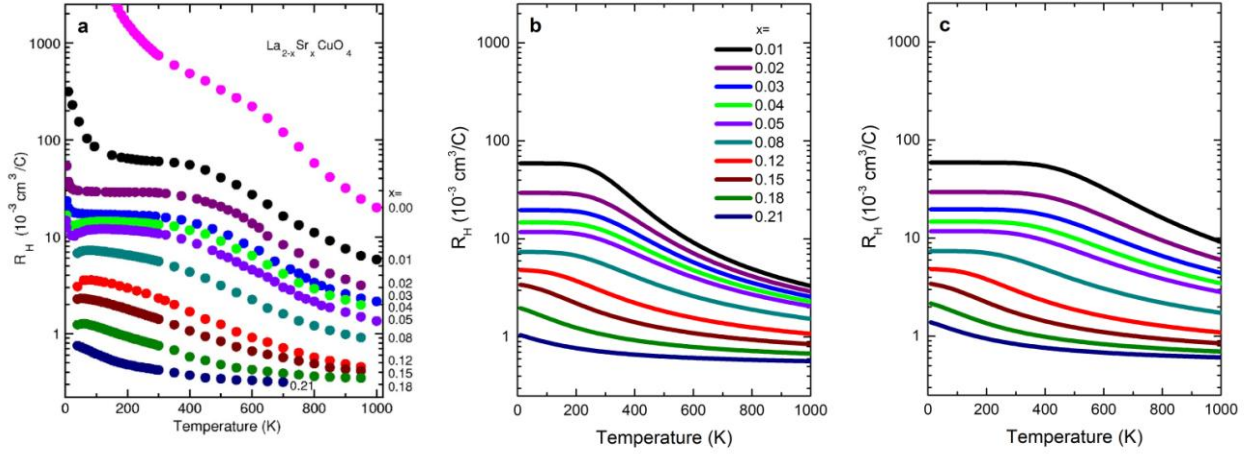

**Figure S6 | Temperature and doping dependence of Hall constant for LSCO. a,** Hall constant for LSCO up to high temperatures (adapted from ref. 30). The low-temperature upturn is due to sample-specific disorder/phase separation effects. **b,** Values calculated from our simple model with gap distribution width 800 K, assuming a doping-independent Fermi surface shape and a linear dependence of the mean gap energy on doping, similar to the phase diagram shown in Fig. 1 in the main text. In order to obtain numerical values for the Hall constant, the size of the LSCO unit cell was taken into account. $R_H$ in experiment levels off at somewhat higher temperatures than in our calculation, possibly as a result of a stronger increase of the mean gap energy in approaching zero doping in the experiment. **c,** Calculated Hall constant with a modified dependence of the mean gap on doping, Eq. (4). The behavior at the lowest doping levels is better captured by allowing for a curvature in the doping dependence of the mean gap. Importantly, our model captures the smooth decrease in optimally doped and overdoped compounds with good precision.

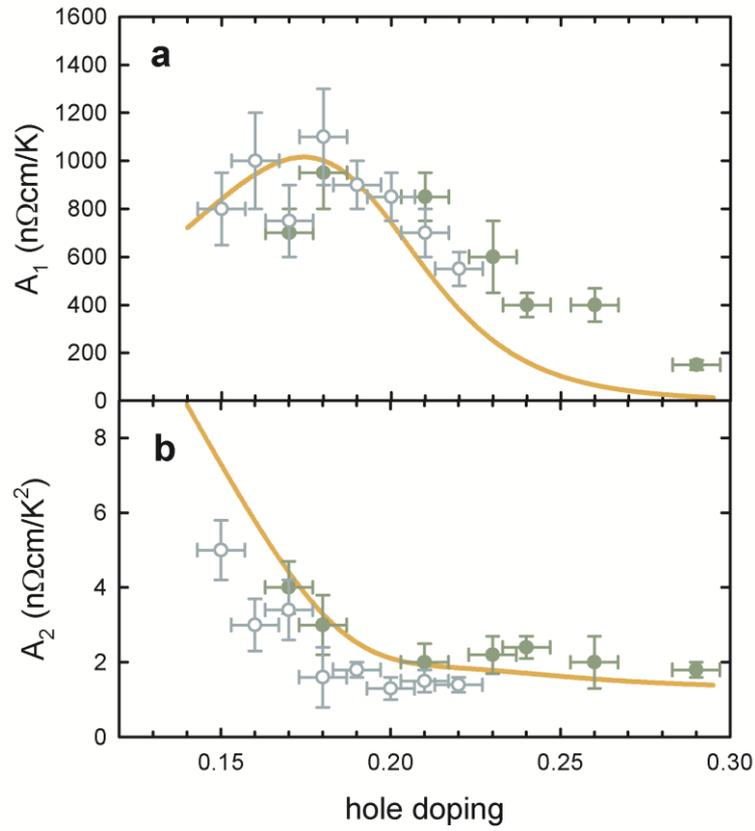

**Figure S7 | Doping dependence of linear and quadratic resistivity coefficients of LSCO. a,** Linear coefficient $A_1(p)$. **B,** Quadratic coefficient $A_2(p)$. The data are from Ref. 35 (full circles) and ref. 23 (empty circles), extracted from measurements up to 200 K at all doping levels. The result for our model (line) is obtained via the same analysis as that employed in ref. 35 (see supplementary text).

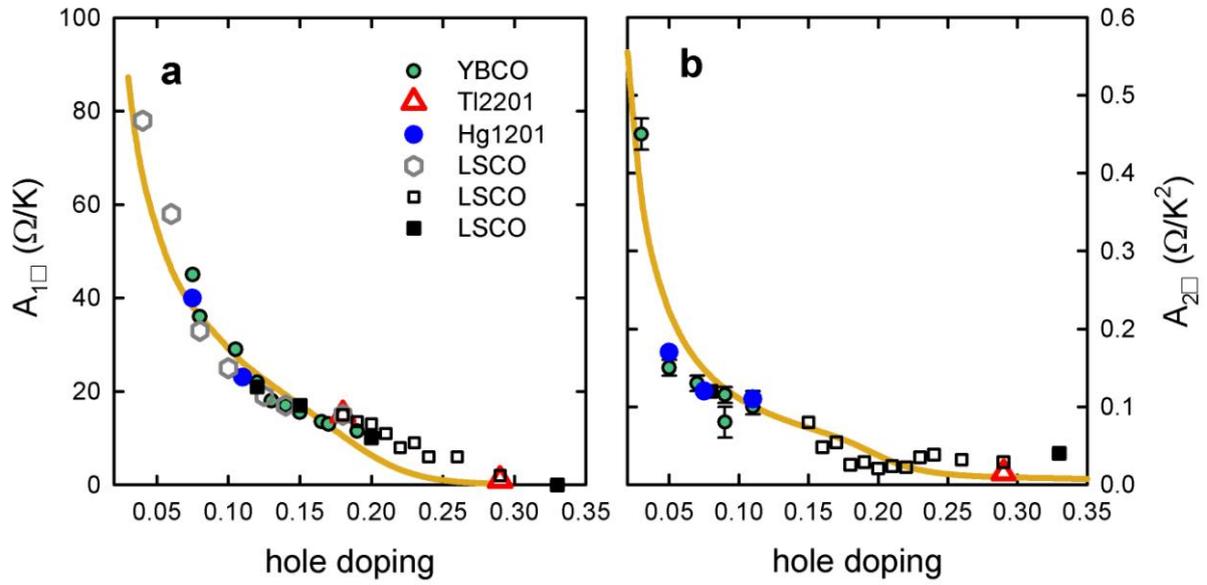

**Figure S8 | Doping dependence of sheet resistance coefficients. a,** Linear sheet resistance coefficient $A_{1\square}$. **b,** Quadratic sheet resistance coefficient $A_{2\square}$. All experimental points are taken from ref. 24 (see references therein for original data sources). For LSCO, data for polycrystalline samples (grey hexagons for $A_{1\square}$) and single crystals (full and empty squares) are shown. The values obtained from our gap distribution model (lines) are calculated as described in the supplementary text.

# Unusual behavior of cuprates

# explained by heterogeneous charge localization

D. Pelc,[1,2] P. Popčević,[3,4] G. Yu,[2] M. Požek,[1*] M. Greven,[2*] and N. Barišić[1,2,3*]

[1]Department of Physics, Faculty of Science, University of Zagreb, Bijenička cesta 32, HR-10000, Zagreb, Croatia

[2]School of Physics and Astronomy, University of Minnesota, Minneapolis, MN 55455, USA

[3]Institute of Solid State Physics, TU Wien, 1040 Vienna, Austria

[4]Institute of Physics, Bijenička cesta 46, HR-10000, Zagreb, Croatia

*Correspondence to: mpozek@phy.hr, greven@umn.edu, neven.barisic@tuwien.ac.at

**Content**

Supplementary Notes

1. Gap distribution
2. Phase diagram and skewness of distribution
3. Phase diagram and choice of doping dependences of distribution parameters
4. Hall coefficient and Hall angle
5. Resistivity coefficients
6. Dome-shaped doping dependence of $T_c$

Figures S1-S8

Supplementary References

**1. Gap distribution.** The cuprates are disordered on several levels, and different experimental probes access different types of disorder. In order to illustrate this salient point, we compile in Fig. S1 some results of local-probe measurements – scanning tunneling microscopy (STM) and nuclear quadrupole resonance (NQR) – for several representative compounds. STM studies of bismuth-based cuprates consistently show a broad distribution of pseudogaps and superconducting gaps[1], and a systematic analysis of different gap scales in $Bi_2Sr_2CaCu_2O_{8+\delta}$ (Bi2212) demonstrated the same underlying disorder on all scales and at all measured doping levels[2]. Copper NQR is a bulk local probe of net electrostatic disorder at the copper sites, which includes doping inhomogeneity, dopant-atom fields, structural distortions, etc. The significant NQR linewidths observed for both $La_{2-x}Sr_xCuO_4$ (LSCO) and $HgBa_2CuO_{4+\delta}$ (Hg1201) demonstrate the presence of intrinsic charge inhomogeneity[3,4] (note that the data for Hg1201 correspond to the quadrupolar satellite in NMR, and the frequencies are referenced to the line midpoint). In $YBa_2Cu_3O_{6+\delta}$ (YBCO), the NQR linewidths are smaller but still substantial (not shown in Fig. S1), despite the fact that this compound is highly crystallographically ordered for some oxygen doping levels[5].

Both electronic and structural origins of intrinsic disorder are possible, and there will always exist an interplay between the two due to electron-lattice coupling. Perovskites in general are prone to structural frustration leading to phase transitions or, if a long-range lattice distortion is avoided, to nano- and mesoscale disorder patterns[6,8]. Such nondiffusive cooperative structural displacements belong to a class of phenomena referred to as martensitic transitions, and are essentially unavoidable in lamellar systems such as the cuprates. The complex ensuing patterns of local unit cell deformations are a plausible candidate for the intrinsic disorder and gap distributions that underlie our phenomenological model. An interesting possibility is that the effects of structural disorder are exacerbated through coupling to strongly-correlated electrons.

Since transport measurements reveal that exactly one hole per $CuO_2$ unit, is (de)localized, it is natural to consider underlying Mott localization as the root mechanism, which is then modified by material complexities. In a simple one-band model without disorder, the (de)localization gaps the entire Fermi surface through a first-order phase transition. Notably, this process does not break any symmetry[7] leaving the Brillouin zone unchanged. However, understanding the evolution of such (de)localization upon doping, in a system with multiple bands (i.e., explicitly including the oxygen degrees of freedom) in the presence of structural/electronic disorder (that presumably causes gradual, inhomogenous (de)localization) creates enormous difficulties for theory. It is therefore not surprising that the 'infrared' physics of the cuprates is extremely complex and exhibits no clear/sharp features in optical conductivity, tunneling spectroscopy and photoemission. Our model packages these complexities in a simple effective manner, without purporting to provide microscopic insight.

The model features only three experimentally-constrained parameters associated with the gap distribution function: $\Delta_0$ (the extrapolated mean gap at zero doping, $\Delta_{p=0}$), $p_c$ (the doping level where the mean gap crosses the Fermi level), and $\delta$ (the width of the gap distribution). In order to calculate the magnitude of the resistivity, the universal scattering rate parameter $C_2$ known from experiment[21] is additionally required. In principle, the gap distribution parameters would be best evidenced from STM, a real-space local probe. Given that the cuprates are inherently inhomogeneous in the bulk[6,8], a surface probe like STM will naturally detect this inhomogeneity as well. An example of the fact that STM provides insight into bulk properties is the remarkably good agreement between STM and bulk X-ray experiments regarding the modulation wave vector of the subtle short-range charge-density-wave (CDW) order in the pseudogap state[9,10,11]. However, only a very limited number of cuprate compounds are accessible to STM, and the gap disorder is expected to be to some extent compound-dependent. STM data are therefore only used as a guide for our calculations, especially on the overdoped side, where there are virtually no relevant data. Extensive measurements on Bi2212 show that the largest local gap scale extracted from STM (denoted $\Delta_1$ in ref. 2) is about 800-1000 K around $p = 0.08$-$0.10$ and that it decreases monotonically with doping. This scale corresponds rather well to the intermediate scale $E_{up}$ of the three scales ($E_{hump} > E_{up} > E_{lp}$) compiled in ref. 12. On the other hand, our gap scale $\Delta_p$ is larger by about a factor of two and better corresponds to the largest scale $E_{hump}$ in ref. 12, which manifests itself in numerous probes[13-18], including optical conductivity[14,15], photoemission spectroscopy[16,17] and tunneling junction experiments[18,19] (Fig. S2b). Figure S2b is a compilation of the doping dependence of these latter results for a number of cuprates. The remarkably good agreement between $\Delta_p$ and $E_{hump}$ implies that the localization of the one hole per $CuO_2$ unit is the highest ranking process, whereas the other energy scales and related processes are emergent.

In order to obtain the generic values of $\Delta_0$, $p_c$ and $\delta$ used in Fig. 1, we simply assume a linear doping dependence of $\Delta_p$, with an extrapolated endpoint approximately equal to $p_c$ and a doping-independent distribution width (see Fig. S2b). Remarkably, in order to achieve detailed quantitative agreement between our model and experiments, these assumptions only need to be slightly relaxed (see below). The generic parameters for the Gaussian gap distribution in Fig. 1 are $\Delta_0$ = 4000 K, $\delta$ = 700 K (Gaussian width) and $p_c$ = 0.2. We note that we express all gap energies in units of Kelvin (with 1 meV ≈ 11.6 K), but that no assumptions related to the physical relationships between energy and temperature scales are made. In particular, in order to calculate transport properties via Eq. (1), only the gap energy values are needed (see also below).

For simplicity, we take $\delta$ to be a constant, in rough agreement with the broad features seen at the $E_{hump}$ scale in photoemission and tunneling experiments and with the scatter in the data in Fig. S2b. Moreover, our choice of $\delta$ is consistent with the STM results on Bi2212, where (lower-energy) gap distribution widths are only weakly doping dependent, and amount to about 500 K (Fig. S1a-c). Since the high-energy 'hump' scale is about a factor of two larger than the largest scale extracted from[2] STM, it is reasonable and in agreement with experimental observations that the width of different gaps scales with their mean gap values, to take the width to be about two times larger as well.

Importantly, our approach is compatible with optical conductivity experiments, which provide another way to estimate the gap distribution parameters. In addition to the coherent Fermi-liquid contribution at low frequencies[20], broad mid-infrared peaks are clearly resolved (see Fig. S2a) that shift to lower frequencies with increasing doping and eventually merge with the coherent response at high doping ($p > 0.20$)[14,15]. This feature was previously identified as a signature of the localized holes[21]. In Fig. S2b, we demonstrate that the mean energy of this peak (detailed doping dependence determined for both LSCO and YBCO[14,15]) indeed coincides with $E^*_{hump}$ and thus $\Delta_p$. In strongly underdoped LSCO ($p$ = 0.02), the width is about 1000 K and the mean energy is approximately 5000 K (see also Fig. S2a).

As shown in Fig. S3, an interesting observation can be made by plotting the temperature and doping dependence of the calculated density of localized carriers along with data for the highest available characteristic temperatures $T^*_{high}$ for LSCO ($T^*_{high} > T^* > T^{**}$) obtained from resistivity and magnetic susceptibility (from ref. 12, where the scale is denoted as $T^*_{hump}$). It can be seen that the data nicely follows the line corresponding to 1 − $p_{loc}$ ~ 20%, which might be an indication of the existence of a critical concentration and underlying percolation process.

**2. Phase diagram and skewness of distribution.** The gap distribution is usually skewed to the high-energy side[1,2,22] (see Fig. S1). In order to quantify and assess the impact of this tendency, we parametrize the distribution as a Gaussian multiplied by its integral:

$$g(\Delta, \Delta_p, \delta) = 2\phi(\Delta, \Delta_p, \delta) \int_{-\infty}^{\alpha\Delta} \phi(\Delta', \Delta_p, \delta) d\Delta' \qquad (1)$$

where α is a dimensionless parameter which controls the skewness of the distribution, and ϕ is a normalized Gaussian distribution with mean $\Delta_p$ and full width at half maximum δ. For α = 0, the distribution reduces to a simple Gaussian, and its skewness can be continuously varied by changing α.

In Fig. S3, we test different values of the skew parameter, as well as different functional forms for the gap distribution, to demonstrate that this yields essentially the same phase diagrams. The main features are rather insensitive to the choice of distribution shape, mainly due to the fact that the resistivity calculation involves an integral over energy. Along with skewed Gaussians, we tested a shifted gamma distribution that features a heavier tail at high energies and gives a broader linear-$T$ resistivity region on the overdoped side of the phase diagram.

**3. Phase diagram and choice of doping dependences of distribution parameters.** The simplest assumption of a linear decrease of the mean energy and doping-independent distribution width may be relaxed by introducing additional parameters and assumptions. While presently such assumptions are purely phenomenological (and detract somewhat from the main message of the simplest possible calculation), we expect a microscopic theory to be capable of providing the true doping dependences, at least in principle. In order to test the robustness of our calculation, we introduce a curvature into the dependence of the mean gap on doping, using a function of the form

$$\Delta_p(p) = \Delta_0 [1 - \tanh(p/p_c)/\tanh(1)] \qquad (2)$$

This function still crosses zero at $p = p_c$ (and contains no additional parameters), but has an upward curvature at higher doping.[23]. A similar function can be used for the dependence of the distribution width on doping, but with a more general form

$$\delta(p) = \delta_0 [1 - \beta \tanh(p/p_c)] \qquad (3)$$

where β is a numerical constant. The more constrained form with β = 1/tanh(1) cannot be used for δ, since it would lead to a zero distribution width at $p_c$ and nonphysical divergences in the calculations. We choose β ~ 0.4, but again it turns out that the exact value is not very important. The main features of the phase diagram are unchanged, but introducing these nontrivial doping dependences of $\Delta_p$ and δ broadens somewhat the region of the phase diagram where the resistivity is linear in temperature on the overdoped side (Fig. S5). In order to introduce also a curvature in $\Delta_p(p)$ at low doping, in line with some experiments (Fig. S2b), we further modify Eq. (2) and cast it in the form

$$\Delta_p(p) = \Delta_0 \left[1 - \left(\tanh(p/p_c)/\tanh(1)\right)^{1/2}\right] \quad (4)$$

without introducing additional free parameters. This form gives better agreement between calculated and measured Hall coefficients for strongly underdoped LSCO (see next Section), as well as a better match between the modeled mean energy (dashed line in Fig. S2b) and the characteristic high-energy scale from experiment. Moreover, it has the physically appealing feature of $\Delta_0$ being approximately 1 eV, the charge transfer gap at zero doping as determined from Hall-effect measurements[30]. Yet again it introduces no considerable changes in the overall picture, and thus in the main text we keep the simplest dependencies for $\Delta_p$ and δ.

**4. Hall coefficient and Hall angle.** Numerous interesting interpretations of the transport behaviour of the cuprates were proposed in the past. However, certain experimental facts were unknown until rather recently. In particular, only in recent years it was firmly established that the charge carriers exhibit universal Fermi-liquid behavior within the pseudogap regime[20,21,24,25] below a characteristic temperature $T^{**}$ ($T^{**} < T^*$): the planar resistivity scales as[24] $\rho \propto T^2/p$, the optical scattering rate exhibits Fermi-liquid temperature-frequency scaling[20], the magnetoresistivity obeys Kohler's rule (Fermi-liquid scaling)[25]. Moreover, the temperature dependence of the cotangent of the Hall angle (i.e., the inverse Hall mobility) is universally $\cot(\Theta_H) \propto C_2 T^2$, not just at temperatures below $T^{**}$, but throughout the whole accessible doping range[21]. These scaling results pose a very strong constraint, as they demonstrate that non-Fermi liquid models[26,27] are not appropriate in the pseudogap and 'strange metal' regimes.

At temperatures below $T^{**}$ (for underdoped compounds), the Hall constant is a rather accurate measure of the doped carrier density[21,30] $p$. It has been long known that the Hall constant exhibits strong temperature dependence above the pseudogap temperature[28,29,30] which, when interpreted in the simplest possible manner, indicates that the effective carrier density acquires a temperature dependence. Although such behavior is not exhibited by textbook Landau Fermi liquids, it is consistent with a repopulation of the Fermi surface from arcs (for $T < T^{**}$) to a large $(1 + p)$ Fermi surface (at high temperatures and doping levels). Indeed, measurements of the Hall constant in LSCO up to 1000 K show a strong decrease, consistent with considerable delocalization of charge[30]. In fact, early work on LSCO already indicated approximate $1/T$ scaling at temperatures[28] above $T^*$, and there were considerations that included a temperature-dependent effective carrier density. In particular, ref. 31 describes a $T^2$ Hall mobility and temperature-dependent carrier density in thallium cuprates; ref. 32 discusses the possibility of a linear-$T$ resistivity arising from a $1/T$ carrier density temperature dependence, and in refs. 33 and 30 the effective carrier number was modeled by including an activation term with a single gap energy.

Yet as remarked in the main text, only recent results for the cotangent of the Hall angle show simple universal quadratic temperature dependence across $T^*$ and to high doping levels, which demonstrates that even in the strange-metal regime the underlying scattering rate is Fermi-liquid-like[21]. The states on the arcs (elongated with increasing temperature and doping) are therefore exactly the same as those found at high doping levels, where clearly a Fermi-liquid exists. In the present work, we therefore keep the experimentally established Fermi-liquid scattering rate throughout.

In addition to enabling a straightforward calculation of the doping and temperature dependences of the resistivity (apart from a non-universal, compound-specific "residual" resistivity contribution[34], our model gives the Hall constant simply as the inverse effective number of carriers $1/p_{\text{eff}}$. In Fig. S6, we compare our result to the data of Ref. 30. The temperature dependences are captured well, even in the optimally-doped region where single-gap fitting is not satisfactory[30]. However, as expected, the absolute magnitudes do not agree completely and require a multiplicative correction as a consequence of the change of Fermi-surface curvature with doping. In LSCO, the curvature decreases toward the antinodal region of the Brillouin zone that is increasingly relevant near optimal doping, which leads to an overall decrease of the Hall constant[21]. In the strongly underdoped compounds, even better agreement between the calculated Hall constant and experiment is obtained by using Eq. (4) for the dependence of the mean gap on doping, which possesses an upward curvature as doping tends to zero (which is not the case for the similar Eq. (2)).

**5. Resistivity coefficients.** Here we compare our model to two other pivotal experimental results. The relatively wide doping range in which the planar resistivity of optimally/overdoped LSCO exhibits linear-$T$ dependence was taken as a sign of 'anomalous criticality'[35]. Whereas this result cannot be easily understood with conventional theories of quantum criticality[35], it naturally follows from our model. Namely, the (de)localization gap disorder causes a rather smooth evolution from the underdoped Fermi liquid, with $p$ mobile carriers at low temperatures, to the overdoped Fermi liquid with $1 + p$ mobile carriers. As the gap structure crosses the Fermi level, an extended region of doping with appreciable linear-$T$ resistivity appears. For a quantitative comparison to the data, we employ the same analysis as in ref. 35 on the resistivity obtained from our model. Note that the model does not take into account the changing geometry of the LSCO Fermi surface with overdoping[36]. Following ref. 35, the model results are fit to a parallel resistor formula $1/\rho = 1/(A_0 + A_1 T + A_2 T^2) + 1/\rho_0$, where $A_0$, $A_1$ and $A_2$ are the resistivity coefficients, and $\rho_0$ a fixed limiting resistivity value corresponding to high-temperature resistivity saturation[35]. The temperature range of the fit is 5 to 200 K at all doping levels, which is arbitrary, but the same as in ref. 35. We again use Eq. (1) from the main text for the effective density of mobile carriers $p_{\text{eff}}$, and we calculate $\rho = C_2 T^2 / p_{\text{eff}}$. The constant $C_2$ captures the universal scattering rate (as deduced from Hall-angle measurements[21]) and determines the absolute value of $\rho$. The comparison of the calculated coefficients $A_1$ and $A_2$ with experiment in Fig. S7 demonstrates remarkable agreement considering the complex Fermi surface of LSCO and the simplicity of our model (e.g., the linear doping dependence of the mean gap $\Delta_p$ beyond optimal doping is probably overly simplistic; a relationship like Eq. (2) would extend the linear-$T$ region to higher doping levels).

We also establish that the maximum in $A_1$ observed experimentally around $p = 0.18$ is a consequence of the limited temperature range of the data and fit (up to 200 K); calculations for a significantly wider range smooth out the maximum (see below). If the upper temperature bound is decreased from 200 K, the maximum becomes even more prominent and its width decreases – this is expected, since the linear-$T$ term appears at low temperatures only close to the doping level $p_c$ where the mean of the gap distribution crosses the Fermi level. Note that $p_c$ derived from an extrapolation of $T^*$ is slightly smaller than our $p_c$ – this is due to the fact that the $T^*(p)$ line in the phase diagram corresponds approximately to 0.9 localized holes per unit cell, while the mean gap $\Delta_p$ corresponds to 0.5 localized holes. The $T^*$ line will thus cross zero at lower doping than the $\Delta_p$ line.

The second experimental study we discuss here is ref. 24. This work analyzes results for numerous cuprates for the planar resistivity above $T^*$ and below $T^{**}$ and concludes for underdoped compounds that the sheet resistance coefficients $A_{1\square}$ ($T$–linear behavior; fit to data for $T > T^*$) and $A_{2\square}$ (from $T^2$ Fermi-liquid fit for $T < T^{**}$) are universal and scale approximately as $1/p$. It was also found that, beyond optimal doping, once superconductivity is suppressed with a large $c$–axis magnetic field and the planar resistivity is best fit to a polynomial form rather than pure linear or quadratic behavior, $A_{1\square}$ decreases sharply whereas $A_{2\square}$ levels off (Fig. S8). We calculate the resistivity coefficients by fitting a polynomial temperature dependence in the low-temperature range (1 – 100 K) to extract $A_{2\square}$. To obtain $A_{1\square}$, we fit a linear function to the calculated resistivity in a temperature range $T^* < T < T^* + 200$ K (where the resistivity is predominantly linear) for underdoped compounds, and above ~0.12 hole doping we fit a parallel resistor formula (as described above) in the temperature range 100 – 400 K. The two procedures to extract $A_{1\square}$ merge smoothly around $p = 0.12$. The extracted model values of $A_{1\square}$ are somewhat sensitive to the fit procedure, as also exemplified by the maximum in Fig. S7. Therefore, $A_{1\square}$ had to be multiplied by a factor of 1.2 in order to best match the experimental values. Note also that photoemission experiments in overdoped LSCO find a Fermi surface with both electron-like and hole-like sections, corresponding to a net carrier number[36] of $1 - p$. For the sake of simplicity, we have neglected any Fermi surface changes and assumed a carrier density of $1 + p$ at high doping (as seen in other cuprates with simple Fermi surfaces[37]). This difference in carrier density may explain the above-mentioned factor 1.2 discrepancy. For the description of the superfluid density of ovedoped LSCO (see Fig. 3), these corrections are unimportant, since only $p_{\text{loc}}$ is needed and the contribution from itinerant carriers is described by Homes' law (see Eq. (2) in the main text).

As seen from Fig. S8, the experimental trends of $A_{1\square}(p)$ and $A_{2\square}(p)$ are well captured by our model. The $1/p$ dependences below optimal doping follow because for underdoped compositions, $p_{\text{eff}} = p$ in a relatively wide temperature region; the simple Fermi-liquid formula for the conductivity then yields $A_{2\square}$ directly, whereas $A_{1\square}$ is the lowest-order correction to the Fermi-liquid dependence in this doping range, and thus also scales with $1/p$. As optimal doping is approached, the complex gap structure begins to cross the Fermi level, and $p_{\text{eff}}$ departs from $p$ (as seen in Fig. 1b). Beyond optimal doping, $A_{1\square}$ strongly decreases as the overdoped Fermi-liquid regime is approached, whereas $A_{2\square} \propto 1/(1+p)$, which is only weakly $p$-dependent. Due to the relatively large temperature range of the fits to the result generated by our model (up to 400 K, compared to 200 K in Fig. S7a), $A_{1\square}$ no longer exhibits a maximum.

In order to best match the LSCO data, the coefficients in Figs. S7 and S8 are obtained with slightly different model parameters than the phase diagrams in Figs. 1, 2, S1 and S2: $\Delta_0 = 3900$ K, $\delta = 800$ K, and $p_c = 0.22$. In LSCO, the extrapolated pseudogap endpoint is indeed slightly shifted to higher doping compared to Hg1201 or YBCO, justifying the employed shift. Given the simplicity of our model, the agreement with experiment is remarkable.

Very recent work attributes the linear-$T$ resistivity to strong quasi-one-dimensional umklapp scattering of antinodal quasiparticles, which do not contribute to the Hall conductance that stems from the nodal (Fermi-liquid) parts of the Fermi surface[38]. It is not obvious how to reconcile this scenario with the experimental fact that $\cot(\Theta_H) \propto T^2$ across $T^{**}$ and $T^*$, and the success of our minimal model in describing many experimental results demonstrates that it is not necessary to invoke such a mechanism. Yet if such a scattering process does play a role, we emphasize that gap disorder is an essential ingredient required to understand the smooth crossover to the overdoped regime.

Our calculations demonstrate that an extended linear-$T$ regime in the resistivity can be obtained in a simple manner, without invoking bad-metal transport or quantum criticality. However, we note that a quantum critical point associated with the emergent unconventional magnetism below $T^*$ may well be present, but without a direct influence on transport and superconducting properties, especially at high temperatures.

**6. Dome-shaped doping dependence of $T_c$.** In conventional superconductors, the effective phonon-induced interaction between quasiparticles is not instantaneous, but occurs on a characteristic timescale related to the phonons involved in the pairing. This gives rise to a relation between the density of electrons and $T_c$, which we believe is also important in the case of the cuprates (regardless of the nature of the pairing mechanism). Within the well-known Eliashberg theory of strongly-coupled phononic superconductors, optic phonons with a characteristic energy $\omega_p$ are the source of the superconducting glue. The interaction is thus relatively confined in space and time, and two quasiparticles must therefore be within a characteristic volume for pairing to occur. The superconducting transition temperature was shown to be given by[39]

$$\ln\frac{T_c}{\omega_p} \approx -\frac{1+\lambda}{\lambda - \mu^* - \lambda\mu^*} \qquad (4)$$

where λ is the electron-phonon coupling and μ* is the Morel-Anderson Coulomb pseudopotential[40] which measures the repulsive electrostatic interaction. The coupling may be obtained, e.g., in a simple pseudopotential approximation (for a nearly-free electron metal), using the adiabatic approximation and effective electron-ion potential $V_q$, which depends on the wave-vector $q$. When the retarded nature of the interaction is taken into account, the coupling becomes[39]

$$\lambda = \frac{1}{r_s} \frac{\langle V^2 \rangle}{\langle \tilde{\omega}_p^2 \rangle} \qquad (5)$$

where $r_s$ is the mean radius of a sphere containing one quasiparticle, and the angular brackets denote averages over all wave vectors. For an isotropic material, $1/r_s$ is clearly proportional to $p^{1/3}$, where $p$ is the density of mobile charge carriers. Hence the effective coupling (and thus the $T_c$) depends on the electron density.

The above example of a phonon-induced retarded interaction showcases an important point that is not restricted to the particular pairing mechanism: a pairing interaction that is local in both space and time induces a dependence of $T_c$ on the density of mobile charge carriers. We have shown compelling evidence that the superconducting glue in the cuprates is closely related to the localized hole. Therefore, one would expect the localization scale $\Delta_p$ to be the relevant energy scale for pairing, equivalent to $\omega_p$ above. However, this is clearly not the only factor that affects the superconducting transition temperature, since otherwise $T_c(p)$ would simply monotonically decrease with increasing doping. The dome-shaped doping dependence of $T_c$ can nevertheless be understood quite generally, without knowing the exact nature of the pairing glue, if one takes into account the fact that the pairing interaction is local and retarded. Recent ultrafast pump-probe experiments have shown that the interaction is retarded, and that the characteristic timescale τ is on the order of 20 fs near optimal doping[41]. This value corresponds roughly to $1/\hbar\Delta_p$. According to our results, the pairing is local, and two mobile holes have to be within the pairing volume on a timescale comparable with τ. On the underdoped side, the mobile-hole density is low, the probability of such an event is diminished, and the pairing energy decreases. A nonzero hole concentration is then needed for pairing because of competing repulsive electrostatic interactions, in analogy with the well-known phonon case above. On the overdoped side, the density is high, but the characteristic energy $\Delta_p$ decreases, again lowering $T_c$. Thus arises the superconducting dome. In the case of the cuprates, the calculation of $T_c$ ought to be modified even if a local-phonon mechanism were responsible for superconductivity. The energies $\Delta_p$ are very high and approach the Fermi energy at low doping, which implies that the adiabatic approximation (or, equivalently, the Migdal theorem in phononic superconductivity[42]) is not satisfied. This is again a more general remark, not necessarily limited to a phonon-related glue. Yet we note that the strong-coupling theory involves optic phonons and an electron-phonon interaction that is local in space, and might thus be of more direct quantitative relevance to the cuprates than might seem at first sight.

Furthermore, since the characteristic energy is spatially inhomogeneous in the cuprates, the mobile holes might self-organize to increase the probability of coupling to the most favorable local sites. The resulting superconducting gap disorder is then detected via superconducting percolation[43,44] in the vicinity of $T_c$, but its relation to the distribution of $\Delta_p$ is likely nontrivial. Self-organization might be the reason why the width of the superconducting gap distribution $T_0$ is de facto universal[43,44], whereas the high-energy gap disorder may show some variation among cuprate families – e.g., the transport data for Hg1201 and LSCO are best described with gap distribution widths differing by about 25% (see main text). Yet the localization gap distributions are still remarkably similar. At low superfluid densities, the overall phase stiffness of the condensate is small and the phase may locally adapt to avoid regions of lower pairing energy. Such superconducting phase relaxation may play an important role[45], with possible modifications to our simple calculation.

Finally, in our model, the pairing glue is provided by the localized hole. As noted in the main text, a two-component electron picture may lead to pairing mediated by localized charge fluctuations. In this respect, several experiments give evidence of the importance of oxygen-oxygen charge transfer within the unit cell. The low temperature tetragonal (LTT) phase in lanthanum-based cuprates involves a small structural change which barely affects normal state transport, but breaks the symmetry of the in-plane oxygens and sharply decreases[46] $T_c$. A similar oxygen symmetry breaking and $T_c$ suppression occurs locally around Zn impurities in copper-oxygen planes, as shown by $^{67}$Zn NQR measurements[47]. The oxygen symmetry is also spontaneously broken by charge density waves[48] which are in competition with superconductivity. A fluctuating oxygen-copper-oxygen polarization could thus play an important role in providing the glue[49]. Experiments that directly probe the intra-unit-cell physics are needed to elucidate the exact superconducting mechanism as well as how exactly the hole gets localized.

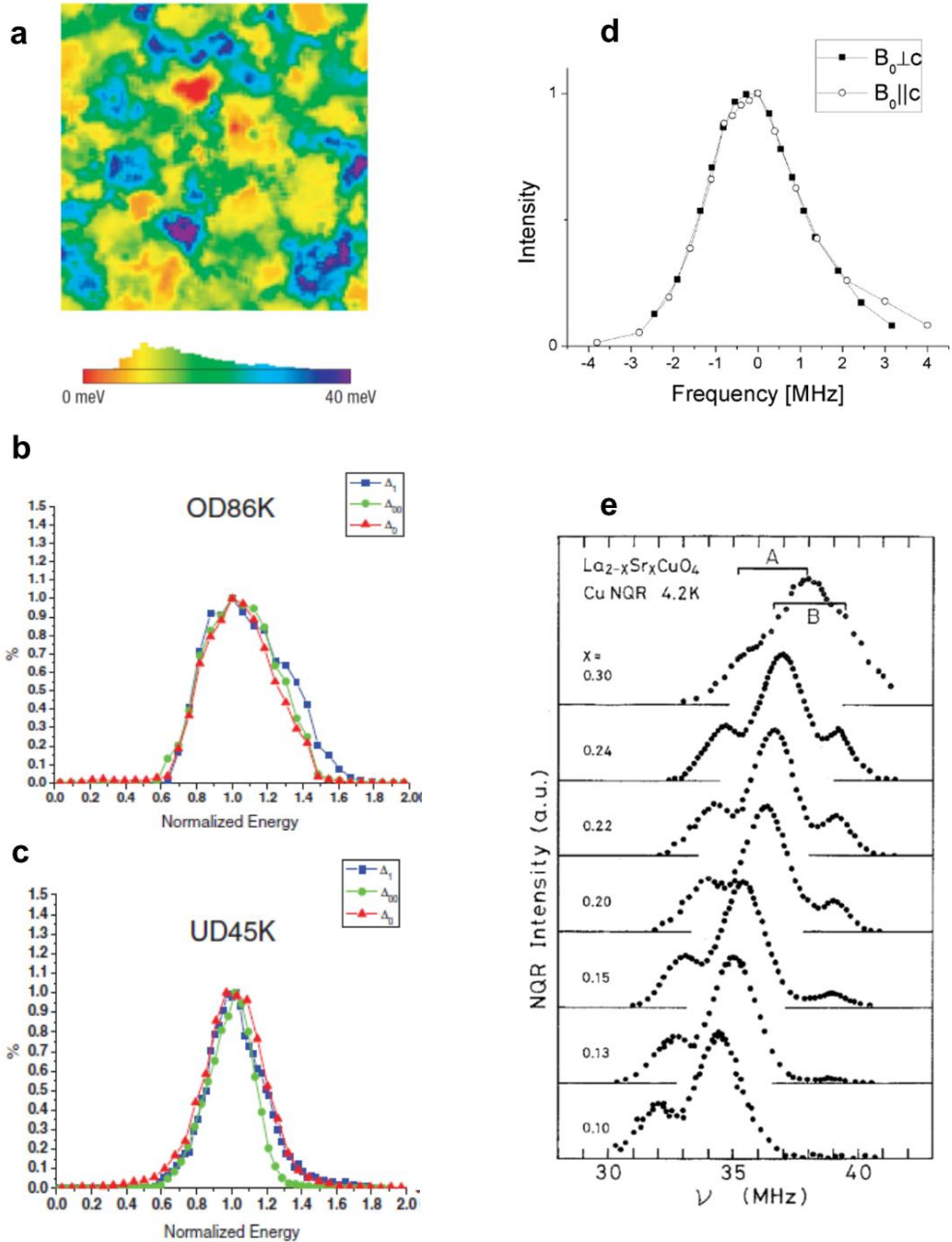

**Figure S1 | Local probes of disorder in cuprates. a,** High-energy gap map from an STM study of Bi2201 and corresponding histogram, in a field of view of 180x180 Å$^2$ (adapted from ref. 1). **b,** and **c,** are distributions of three fitted gap components from an STM study of Bi2212 for an overdoped and underdoped sample, respectively (adapted from ref. 2). Significantly, the same underlying inhomogeneity manifests itself in all gap scales. The energies are normalized to the mean values, but the distribution widths on an absolute energy scale are essentially the same. **d,** $^{63}$Cu NMR quadrupolar satellite line in optimally-doped Hg1201, where the linewidth is a measure of local electrostatic disorder (adapted from ref. 3). **e,** Doping dependence of nuclear quadrupole resonance (NQR) spectra of LSCO in a wide doping range (adapted from ref. 4), showing that the line-width is essentially unchanged across the phase diagram. Notably, widths of the $^{63}$Cu A lines are slightly larger than for Hg1201. The smaller feature below the $^{63}$Cu A line is the $^{65}$Cu A line, whereas the B line stems from sites close to the Sr impurities.

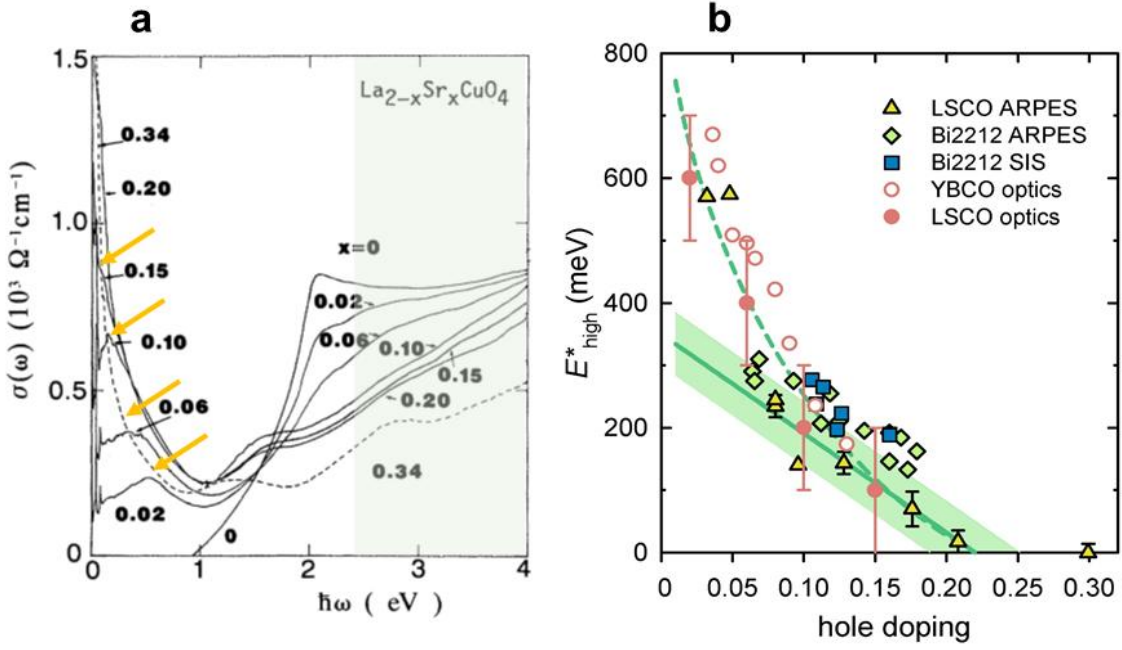

**Figure S2 | High-energy features in cuprates. a,** Optical conductivity of LSCO at 300 K, showing a mid-infrared feature at energies consistent with our mean delocalization gap (arrows indicate the characteristic feature that corresponds to the mean gap value) evolving from the charge transfer gap and merging into the coherent Drude peak at high doping. Adapted from ref. 14. **b,** Comparison of the highest-energy characteristic scale in different cuprates. The ARPES and SIS data correspond to the 'hump' scale, while the optical conductivity data are the energies of the mid-infrared peak. The solid green line is our parameterization of the localization gap for LSCO, whereas the shaded green band indicates the gap distribution width. The dashed line is an alternative parameterization, Eq. (4), which includes upward curvature and extrapolates to the charge transfer gap (~ 1 eV) at zero doping. ARPES and SIS data are adapted from ref. 12while the data are from multiple experiments, see ref. 12 for original references. LSCO optical conductivity peak energies are from **a**, and YBCO optical conductivity is adapted from ref. 15.

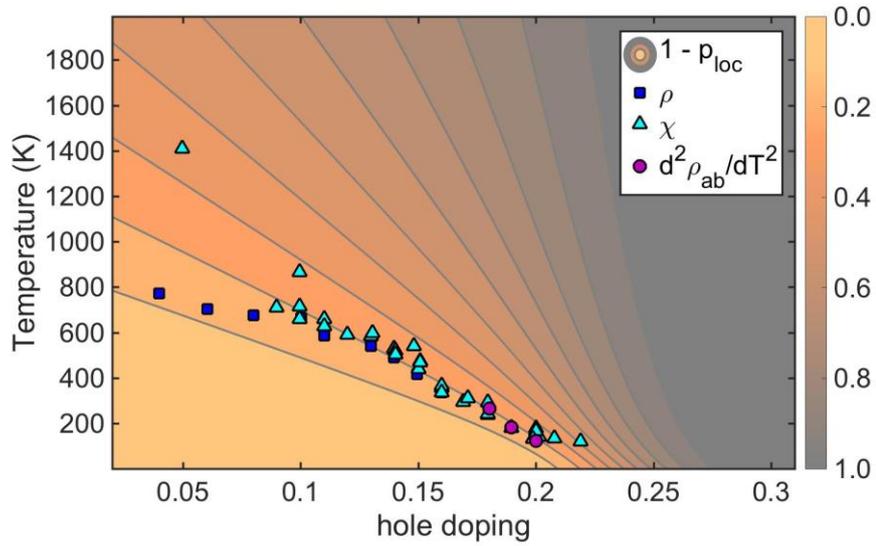

**Figure S3 | Characteristic temperature and localization.** Comparison for LSCO of the highest measured characteristic temperature scale $T^*_{high}$ from resistivity and magnetic susceptibility measurements with the calculated fraction of localized holes. The iso-lines on the contour plot correspond to increments of 10%. It is seen that, between $p \sim 0.10$ and $0.20$, the data closely follow the line for 20% delocalization. This may hint at an underlying percolation/connectivity transition. Data are from ref. 12 and references therein.

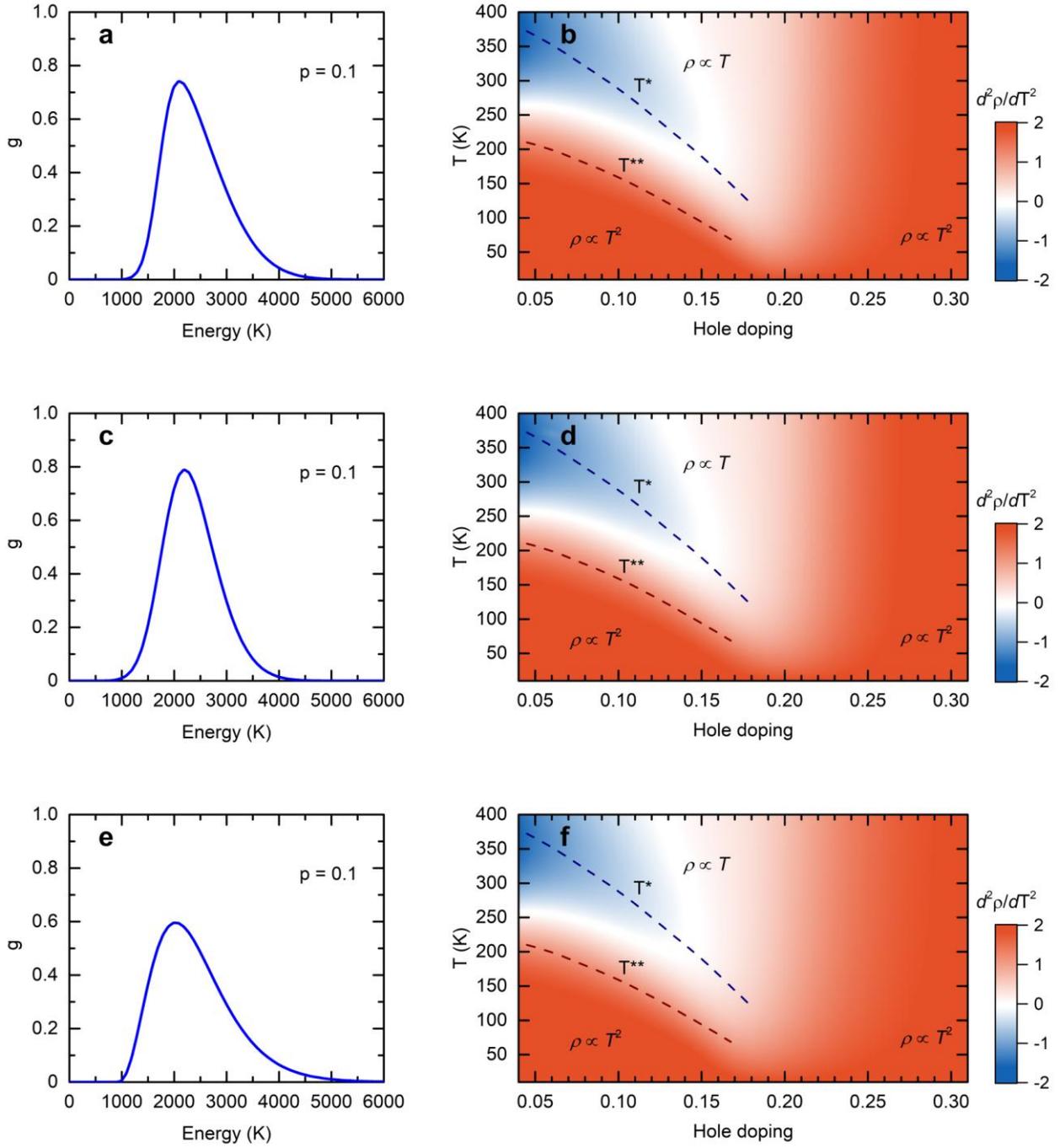

**Figure S4 | Normal state phase diagram for different gap distributions. a,** Skewed Gaussian distribution, with skew parameter α = 4 and width parameter δ = 1000 K. **b,** The corresponding resistivity curvature phase diagram. **c,** A less skewed Gaussian distribution, with skew parameter α = 1 and width parameter δ = 800 K. **d,** The corresponding phase diagram. **e,** A shifted gamma distribution, with exponent $l$ = 3 and width parameter δ = 400 K. **f,** The corresponding phase diagram. All distributions are shown at nominal doping level $p$ = 0.10, and a linear dependence of the mean gap energy on doping is assumed. The distributions are normalized to the area under the curves.

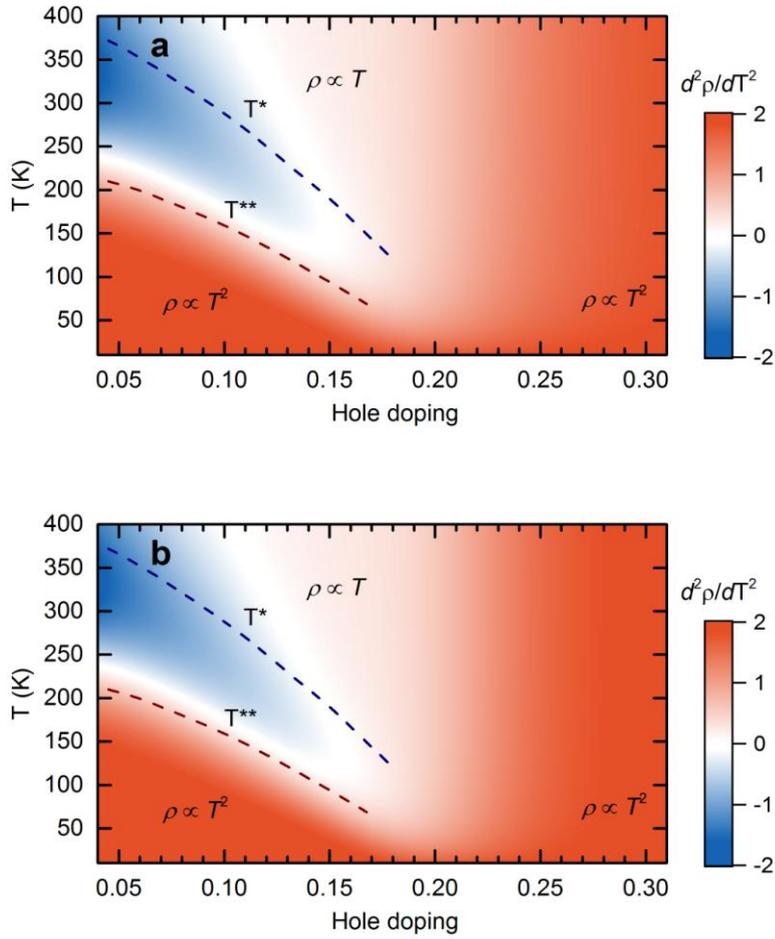

**Figure S5 | Normal-state phase diagrams for two doping dependences of the gap distribution parameters.** Hyperbolic tangent doping dependence of the distribution width and mean, with two values of the width doping dependence parameter (see supplementary text): **a,** $\beta = 0.4$, and **b,** $\beta = 0.6$.

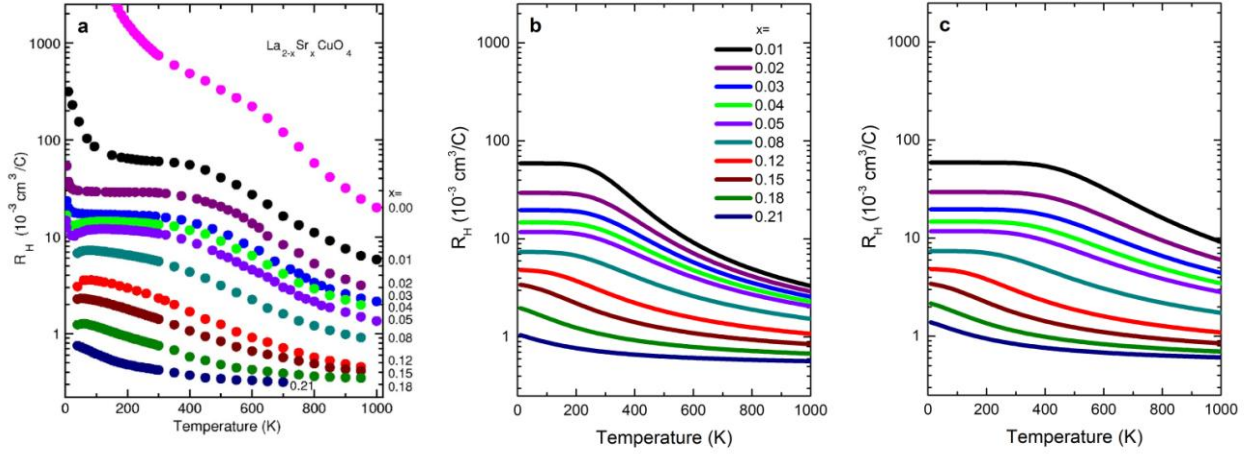

**Figure S6 | Temperature and doping dependence of Hall constant for LSCO. a,** Hall constant for LSCO up to high temperatures (adapted from ref. 30). The low-temperature upturn is due to sample-specific disorder/phase separation effects. **b,** Values calculated from our simple model with gap distribution width 800 K, assuming a doping-independent Fermi surface shape and a linear dependence of the mean gap energy on doping, similar to the phase diagram shown in Fig. 1 in the main text. In order to obtain numerical values for the Hall constant, the size of the LSCO unit cell was taken into account. $R_H$ in experiment levels off at somewhat higher temperatures than in our calculation, possibly as a result of a stronger increase of the mean gap energy in approaching zero doping in the experiment. **c,** Calculated Hall constant with a modified dependence of the mean gap on doping, Eq. (4). The behavior at the lowest doping levels is better captured by allowing for a curvature in the doping dependence of the mean gap. Importantly, our model captures the smooth decrease in optimally doped and overdoped compounds with good precision.

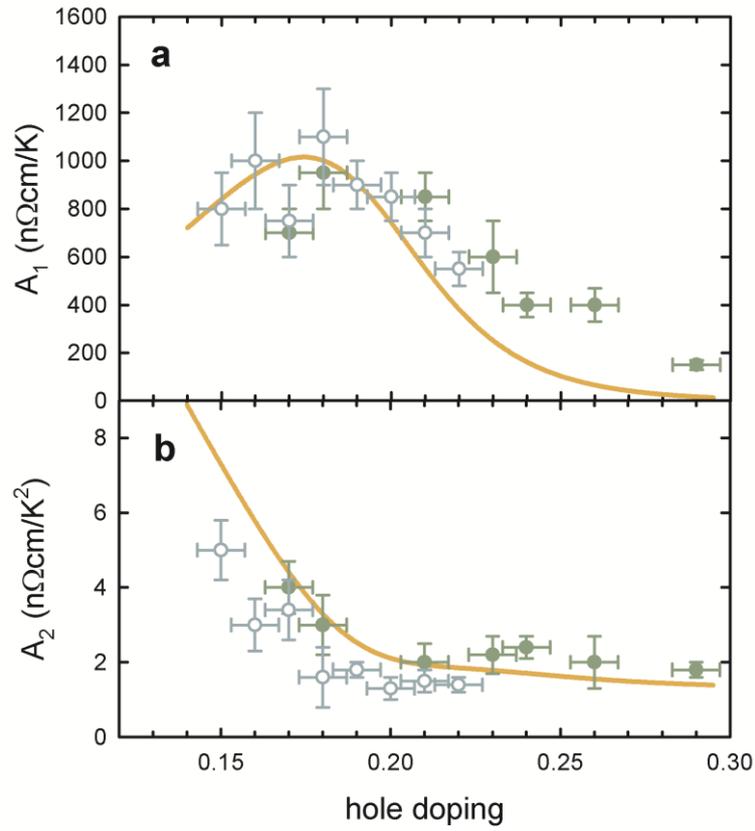

**Figure S7 | Doping dependence of linear and quadratic resistivity coefficients of LSCO. a,** Linear coefficient $A_1(p)$. **B,** Quadratic coefficient $A_2(p)$. The data are from Ref. 35 (full circles) and ref. 23 (empty circles), extracted from measurements up to 200 K at all doping levels. The result for our model (line) is obtained via the same analysis as that employed in ref. 35 (see supplementary text).

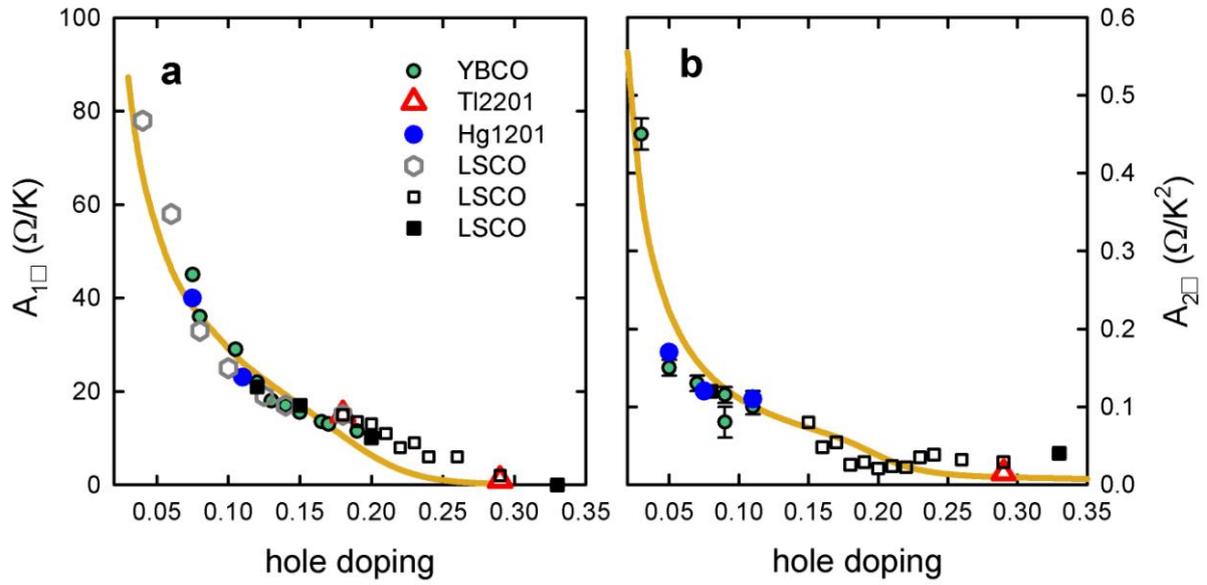

**Figure S8 | Doping dependence of sheet resistance coefficients. a,** Linear sheet resistance coefficient $A_{1\square}$. **b,** Quadratic sheet resistance coefficient $A_{2\square}$. All experimental points are taken from ref. 24 (see references therein for original data sources). For LSCO, data for polycrystalline samples (grey hexagons for $A_{1\square}$) and single crystals (full and empty squares) are shown. The values obtained from our gap distribution model (lines) are calculated as described in the supplementary text.